\documentclass[nonblindrev]{informs3_no_remark} 

\newif\ifcomm
\commtrue
\ifcomm
    \newcounter{commentNumberI}
     \newcommand{\Noa}[1]{\addtocounter{commentNumberI}{1}{{({\color{blue} {(\arabic{commentNumberI}.)} Noa: #1})}}} 
     \newcommand{\Andrew}[1]{\addtocounter{commentNumberI}{1}{{({\color{teal} {(\arabic{commentNumberI}.)} Andrew: #1})}}} 
       \newcommand{\Gal}[1]{\addtocounter{commentNumberI}{1}{{({\color{red} {(\arabic{commentNumberI}.)} Gal: #1})}}} 
    \newcommand{\A}[1]{\textbf{[#1]}}       
    
\else
    \newcommand{\A}[1]{}
    \newcommand{\Noa}[1]{}
    \newcommand{\Gal}[1]{}
    \newcommand{\Andrew}[1]{}
\fi

\OneAndAHalfSpacedXI 

\usepackage{natbib}
 \bibpunct[, ]{(}{)}{,}{a}{}{,}%
 \def\bibfont{\small}%

 \usepackage{tikz,calc}
 \usetikzlibrary{decorations.markings}
 \usepackage{soul}
 \setul{}{0.75pt}
 \setuloverlap{1.5pt}

\usepackage{subfigure}
\usepackage{mwe}
\usepackage{lscape,booktabs,longtable}
\usepackage{amsmath}

\usepackage[colorlinks=true,urlcolor=blue,citecolor=blue,pdfstartview=FitH]{hyperref}
\usepackage{breakurl}

\usepackage{comment}
\usepackage{enumerate}
\usepackage[ruled,vlined]{algorithm2e}
\usepackage[graphicx]{realboxes}
\usepackage{multirow}

\newcommand{\tb}{\textcolor{blue}}

\newcommand{\edit}{{}}

\newcommand{\eProof}{\hfill \qed\vspace*{0.2cm}}
\newcommand{\Ex}{\mathbb{E}}

\newcommand{\calB}{\mathcal{B}}
\newcommand{\calC}{\mathcal{C}}

\TheoremsNumberedThrough     
\ECRepeatTheorems

\EquationsNumberedThrough    

\begin{document}

\RUNAUTHOR{}
\RUNTITLE{The Hybrid Hospital}
\TITLE{The Hybrid Hospital: Balancing On-Site and Remote Hospitalization}

\ARTICLEAUTHORS{%
\AUTHOR{Noa Zychlinski$^1$, Gal Mendelson$^1$, Andrew Daw$^2$}
   	\AFF{$^1$ Faculty of Data and Decision Sciences, Technion -- Israel Institute of Technology, Haifa 3200003, Israel\\
    $^2$ Marshall School of Business, University of Southern California, 3670 Trousdale Pkwy, Los Angeles, CA 90089, USA\\
    noazy@technion.ac.il,~~galmen@technion.ac.il,~~dawandre@usc.edu\\
    Aug 22, 2024}
} 

\ABSTRACT{\textbf{Problem definition:} \edit{Hybrid hospitals offer both on-site hospitalization} and remote hospitalization through telemedicine. These new healthcare models require novel operational policies to balance costs, efficiency, and patient well-being. Our study addresses two \edit{first-order} operational questions: (i) how to direct patient admission and call-in based on individual characteristics and proximity and (ii) how to determine the optimal allocation of medical resources between these two hospitalization options \edit{and among different patient types}. 

\textbf{Methodology/results:} We develop a model that uses Brownian Motion to capture the patient's health evolution during remote/on-site hospitalization \edit{and during travel}. \edit{{Under cost-minimizing}} call-in policies, we find that remote hospitalization can be cost-effective for moderately distant patients, \edit{as the optimal call-in threshold is non-monotonic in the patient's travel time}. \edit{Subject to scarce resources, the optimal solution structure becomes equivalent to a simultaneous, identically sized increase of remote and on-site costs under abundant resources}. 
\edit{When limited resources must be divided among multiple patient types, the optimal thresholds shift in non-obvious ways as resource availability changes. Finally, we develop a practical and efficient policy that allows for swapping an on-site patient with a remote patient when the latter is called-in and sufficient resources are not available to treat both on-site.}

\textbf{Managerial implications:} Contrary to the widely held view that telemedicine can mitigate rural and non-rural healthcare disparities, our research suggests that on-site care may actually be more cost-effective than remote hospitalization for patients in distant locations, due to \edit{(potentially overlooked) risks during patient travel}. This finding may be of particular concern in light of the growing number of ``hospital deserts'' amid recent rural hospital closures, as these communities may in fact not be well-served through at-home care. 
Such insights on cost-effectiveness, proximity, and possible patient deterioration can guide healthcare decision-makers and policymakers in shaping future healthcare delivery and design.}


\maketitle


\section{Introduction}

The COVID-19 pandemic has significantly propelled the adoption of virtual services, with telemedicine now playing a prominent role in the realm of healthcare (\citealp{jnr2020use,kadir2020role}). Telemedicine facilitates the remote delivery of clinical services through real-time communication, connecting patients and healthcare providers via video conferencing and remote monitoring (\citealp{monaghesh2020role}). These virtual services offer several advantages, such as cost savings related to travel and reduced exposure to diseases, which may ultimately enhance the efficiency of healthcare delivery (\citealp{hur2020usefulness}). 

Recent advancements in telemedicine now enable sophisticated remote medical services, including home hospitalization as an alternative to traditional on-site care (\citealp{zychlinski2024tele}). \href{https://beyond-en.sheba.co.il/}{Sheba Beyond}, a pioneering virtual hospital affiliated with Sheba Medical Center and thus ranked among the world's top medical systems by \href{https://www.newsweek.com/worlds-best-hospitals-2022}{Newsweek}, offers remote examination, monitoring, and online rehabilitation programs. 
Their goal is to enhance accessibility to top-tier medical expertise for all prospective patients, aligning with the prediction that remote hospitalization will become a widespread offering among major hospital networks.
Indeed, virtual hospitals are becoming popular across the world, such as in Australia (\citealp{hutchings2021virtual}), China (\citealp{francis2021predictors}), and the United States.
For example, in the US, this trend is well underway: 186 hospitals participated in the ``Acute Hospital Care at Home'' program during its inaugural year (\citealp{clarke2021acute}), which permitted Medicare-certified hospitals to deliver inpatient-level care to patients within the comfort of their homes.  A recent \href{https://www.mckinsey.com/industries/healthcare/our-insights/virtual-hospitals-could-offer-respite-to-overwhelmed-health-systems}{McKinsey \& Company} comprehensive report stated that virtual hospitals have the potential to provide significant relief to overburdened healthcare systems. In particular, they project that virtual hospitals could unlock bed capacity, reduce the need to build new hospitals\edit{,} and save hundreds of millions of dollars (\citealp{boldt2023virtual}). The \href{https://www.aha.org/hospitalathome}{American Hospital Association (AHA)} has similarly promoted the concept through the \emph{Hospital-at-Home} components of their ``Value Initiative" public cost-reduction campaign. 

Often times, such campaigns naturally associate the potential benefits of home hospitalization with rural patients. Frequently referred to as ``hospital deserts'' due to their significant distance from healthcare centers, many rural communities face healthcare accessibility challenges worldwide, affecting millions of individuals in large countries like the United States, China, Brazil, and England (\citealp{behrman2021society,jiao2021identifying,gong2021evaluating,noronha2020covid,verhagen2020mapping}). These under-served areas lack proximity to medical facilities, leading to delays in seeking care, limited access to timely interventions, and increased health risks. Transportation hurdles further complicate the problem, as rural residents must contend with limited options and lengthy journeys, often resulting in worsened health conditions by the time they reach a hospital (\citealp{kelly2014travelling}).
Additionally, as highlighted in a \edit{July 2024} report by the \edit{\href{https://ruralhospitals.chqpr.org/downloads/Rural_Hospitals_at_Risk_of_Closing.pdf}{Center for Healthcare Quality and Payment Reform (CHQPR)}}, \edit{more than 100 rural hospitals have closed in the past decade in the United States, and another 700 rural hospitals, representing more than 30\% of the nation's rural healthcare facilities, face the risk of closure --- for 360 of these hospitals, the risk of closure is categorized as \emph{immediate}}.

This tenuous state of rural healthcare offerings is compounded by growing evidence of a disparity in the health conditions of people in rural areas relative to those who live in non-rural areas \citep{lewis2022people}. For example, in the United States, data from the National Vital Statistics System has shown that in the two decades from 1999 to 2019,  although the overall death rate (deaths per 100,000) has declined, a widening gap has emerged between the rates of death in rural and non-rural communities \citep{curtin2021trends}. Even more troublesome is that this gap is consistent across the 10 leading causes of death, with the widest disparities occurring in the fatality rates for heart disease, cancer, and chronic lower respiratory diseases. Furthermore, these trends are consistent when controlling for demographic factors like age, race, and sex \citep{cross2021rural}. Similarly, data shows that the rate of death from the COVID-19 pandemic in non-metropolitan areas  out-paced the same rate in metropolitan ones \citep{ullrich2021covid}. This heightened deadliness of serious disease in rural communities in the US is coupled with the noted growth of addiction, overdoses, and suicide (so-called ``deaths of despair,''  \citealp{case2015rising,case2017mortality}) and increased mortality of unintentional injuries, such as from traffic and firearms \citep{olaisen2019unintentional}. 

While telemedicine networks have shown promise in improving healthcare in rural areas (\citealp{ishfaq2015bridging}) and could also potentially serve as a viable alternative to mitigate the impact of hospital closures, our paper underscores a critical issue: patients residing in remote areas, who ostensibly stand to gain the most from home hospitalization, also face the highest risks when called-in to the hospital. \edit{Because they inherently must travel greater distances to reach the hospital, remote patients called-in to the hospital may arrive} in deteriorated states, ultimately leading to prolonged and more expensive hospitalizations. 
Hence, hybrid hospitals that offer both on-site and remote hospitalization services present new operational challenges, necessitating the development of innovative models and policies that ensure cost-effectiveness while maintaining the highest standard of patient care. \edit{The heretofore overlooked impact of patient travel in hybrid health networks is the impetus for this paper's first-order, static-planning analysis of hybrid hospitalization.}

More specifically, our study focuses on a hybrid hospital setting that incorporates a virtual Emergency Department (ED), which patients can access when they experience illness. In this model, medical professionals conduct remote examinations and consultations with patients. Subsequently, based on their assessment, doctors decide whether to admit the patient for remote hospitalization or advise immediate travel to the hospital for on-site admission. 
For patients admitted remotely, their physical examination is conducted using telehealth technologies, such as TytoCare\textsuperscript{\tiny\textregistered}, a digital platform specifically designed for remote physical assessments (\citealp{zychlinski2024tele}). During these examinations, both data and visual information are recorded and transmitted to the physician. Then, a summary of the visit is provided, which may include orders for blood tests, medication orders and instructions. Remotely admitted patients have two potential outcomes: recovery with subsequent discharge, or, in the event of health deterioration, they are advised to travel to the hospital for on-site admission and the continuation of their treatment. We refer to this as a ``call-in'' scenario. 

Therefore, the first fundamental question we address is how to optimally set the call-in policy so as to minimize the total operational cost. That is, based on each patient's characteristics, \edit{the hybrid hospital must first} decide whether to admit the patient remotely or on-site. In the former case, one also needs to decide at which health condition to call the patient in to the hospital. The marginal improvement cost in each hospitalization option and the patient's proximity to the hospital and anticipated further deterioration while traveling each play important roles in these decisions. 

The second question we address in this paper is related to the way the hybrid hospital/ward allocates its resources. In Sheba Beyond, the medical staff of each hybrid ward is divided into two teams, each is responsible either for remote or on-site hospitalized patients, which is what we assume throughout this paper. Therefore, the question is how to allocate these resources to these two groups. This decision goes hand in hand with the call-in policy, since the decision on when to call in patients determines the workload for each group.

To address these two questions, we introduce a model that captures \edit{the stochastic progression of} patients' health condition via an acuteness ``score'' that aggregates clinical measurements for supporting discharge decisions. Such scores are common {in} practice. The Aldrete system, for example, is an acuteness score to determine readiness for discharge post-surgery  (\citealp{aldrete1994discharge}); other scores were developed for specific diseases such as pneumonia (e.g., \citealp{capelastegui2008pneumonia}), for cardiac patients (the Anderson-Wilkins acuteness score; \citealp{anderson1992electrocardiographic}), or for patient assessment in SNFs and rehabilitation facilities (e.g., the ADL score; \citealp{bowblis2014medicare}). 
We capture the system's dynamics by modeling the individual evolution of the patient's health condition through remote and on-site hospitalization using Brownian Motions (BMs) whose parameters depend on patients' characteristics. That allows us to capture the fact that patient's health score improves, on average, while being treated, yet may nevertheless deteriorate due to the randomness in recovery across patients. The relevant properties in our analysis are hitting time statistics---averages and probabilities---that determine length of stay (LOS) in both hospitalization options and the call-in likelihood due to deterioration at home.

Our work sheds light on the complex operational aspects in managing hybrid hospitals.
The questions we address in this paper are ones of \textit{design}. Rather than taking the service content at each location as fixed, we optimize it to meet system-level goals by setting the treatment mix of each patient profile as well as the allocation of resources between the two hospitalization locations.

The following are the main contributions of this paper:
\begin{itemize}
\item[\edit{\textbf{Optimal design of hybrid hospitalization:}}] As a modeling contribution, we study the operations, design, and management of an innovative acute-care system comprising both on-site and remote hospitalization. By capturing the random dynamics of patients' health scores and this evolution's dependence on the manner of care, we provide a practical framework for determining the optimal treatment blend and call-in policy based on individual patient characteristics and their travel time to the hospital. We explicitly address critical questions that hinge on two essential factors: (i) the disparity in marginal hospitalization costs between on-site and remote hospitalization, and (ii) the feasibility of call-in as opposed to exclusively remote hospitalization. 

\item[\edit{\textbf{First-principles modeling at the patient level:}}] \edit{To address these questions, we propose 
to model the patient’s score first and then the queueing system second, rather than the other way around.
Central to this paper's pursuit is the previously overlooked impact of patient travel time (or distance) within a healthcare system with remote care offerings. (See the next paragraph.) Brownian motion, with its drifts and diffusion parameters, thus becomes a natural, first-order model for how the system becomes sensitive to patient travel in the potential progression of their clinical severity. This modeling approach offers the flexibility and tractability in mapping service-content decisions to outcomes. If one were to try to create the same type of drift-diffusion structures as the BM model using typical queueing-system-first assumptions, it would require much more restrictive assumptions, such as a state-dependent exponential random variable. By contrast, the BM orients the model around the focal decisions, the call-in threshold and the resulting optimal resource allocation. Hence, our models focus on the individual, rather than on the system.
}


\item[\edit{\textbf{Consequences of patient travel on hybrid healthcare performance:}}]
Both intuition and nascent public policy around home hospitalization suggests that remote hospitalization  and telemedicine could offer a geographic panacea for health outcomes, enhancing healthcare access in remote locales and thus bridging disparities between rural and non-rural areas. Our research, however, demonstrates that, due to the increased risk of deterioration {during} lengthy travel times to the hospital, it could be preferable for the hospital to direct distantly located patients to on-site care. 
\edit{In a series of policy-level insights, we find that, even if the hybrid hospital has unlimited resources,
remote hospitalization will be cost-effective only for those patients who reside up to a moderate distance from the hospital. Specifically, our model identifies that, for patients whose \emph{marginal} treatment costs are higher in the hospital, remote hospitalization is cost-effective only if their distance does not exceed a certain radius; furthermore, this distance will shrink for patients with poorer initial health scores. For patients whose {marginal} treatment costs are higher at home, the optimal call-in threshold is non-monotonic as a function of patient distance, and the range of distances for which remote hospitalization is viable will likewise shrink for poorer initial health scores.}
In light of the broadly-documented evidence of worse baseline health conditions in rural communities, we discuss how our model's insights caution against the prevailing assumption that remote hospitalization would benefit rural patients.

\item[\edit{\textbf{Allocation strategies in hybrid healthcare:}}] 
\edit{To accompany the preceding isolated effects of travel with actionable prescriptions, we also study the setting when} a fixed amount of resources, such as medical professionals, must be distributed between the two modes of care. Three distinct \edit{structural} cases, contingent on the ratio of marginal improvement rates, are characterized within the system's workload and feasibility region: they both  can exhibit an increasing, decreasing, or unimodal pattern in relation to the call-in threshold. Leveraging these findings, we subsequently derive insights into optimal resource allocation. We find that the impact of scarce resources is a simultaneous increase of both remote and on-site cost rates by \textit{the same} value, without altering the solution structure and properties from the case where resources are abundant.  
Notably, we observe that the optimal allocation of resources is non-monotone with the total amount of resources. In some cases, as the total resource pool becomes more limited, a larger proportion may be allocated to one hospitalization option while reducing the allocation to the other. This highlights the dynamic nature of resource allocation in hybrid hospitals.

\item[\edit{\textbf{Management for hybrid health networks across heterogeneous populations:}}] 
\edit{When there are multiple types of patients that ``compete'' for limited resources, the optimal thresholds and allocation of workload changes in non-obvious ways as the amount of resources changes. Therefore, the call-in policies must adapt to resource availability, with call-in thresholds adjusting to ensure cost-effective patient care. Furthermore, understanding the effect of different patient's characteristics can help healthcare providers optimize treatment strategies, ultimately improving patient outcomes and resource utilization.}
\edit{Beyond the optimal system design, we utilize the model components to propose a practical dynamic policy that enables swapping an on-site patient with a remote patient when the latter is called in and on-site resources are insufficient to treat them. Through simulations across various scenarios, we demonstrate that the suggested policy outperforms an intuitive policy which chooses to swap the the patient with the current best health condition. }

\end{itemize}

\vspace{2mm}

The rest of the paper is organized as follows. Section \ref{sec: Literature Review} includes a brief review of the related literature.
In Section \ref{sec: The model (new)}, we introduce our model and the
optimization problem. In Section \ref{sec: Preliminary Results}, we include  preliminary analyses on the system's workload and feasibility region. \edit{For a homogeneous patient population,} the main results of the paper are presented in Section \ref{sec: Results}. \edit{In Section \ref{sec: Multiple Item Types}, we extend the results to accommodate multiple types of patients competing over limited resources. Section~\ref{subsec: A Dynamic Policy} then conducts simulation experiments to demonstrate how this modeling framework can be used for real-time decisions, namely through the dynamic swap policy.} Lastly, in Section \ref{sec:Conclusions and Direction for Future Research}, we provide some concluding remarks and suggest a few directions for future research.  All
proofs appear in the appendix.

\section{Literature Review}
\label{sec: Literature Review} 

This paper is related to two main lines of literature. The first applies stochastic modeling to study the operations of health services. The second is related to health progression modeling. We provide here a brief review of the related literature along these two streams.

Stochastic modeling and queueing systems have been used to address many different healthcare applications and derive associated operational insights and policies (e.g., \citealp{mills2013resource,shi2016models}). One of the challenges in managing such complex healthcare systems is how to allocate scarce resources and prioritize patients over these resources  (e.g., \citealp{Ziya:2018}). While classical models in queueing theory assume that service times are independent random variables with fixed, if not identical, distributions, empirical studies show that there is flexibility in setting transfer/discharge decisions in healthcare; these decisions, in turn, have an effect on patient outcomes \edit{and LOS} (\citealp{kc2012econometric,bartel2020should}). Here, we build upon prior works that have shown the benefit of modeling in finer detail, such as the controlled queueing models studied in works like \cite{hopp2007operations} and \cite{chan2014use}.
 
Protocols for adaptive discharge of individual patients, from a single station, were developed in \cite{shi2021timing}, where a Markov decision process (MDP) is integrated with data to support discharge decisions from inpatient wards. They suggested an efficient dynamic heuristic that balances personalized readmission-risk prediction and ward congestion. Perhaps most similar to our setting is \cite{Hematology2021}, which developed discharge rules specifically for hematology patients. For these, a longer hospital stay carries risk (infections) but also the ability to take care of such infections.
\edit{Relative to \cite{shi2021timing}, \cite{Hematology2021}, and the prior literature, we go beyond a single-station analysis to study a new setting: the hybrid hospital, which includes both on-site and remote hospitalization. Our focus includes decisions on patient hospitalization option, call-in thresholds for remote patients, and resource allocation, considering patient characteristics and distance from the hospital. 
Furthermore, to the best of our knowledge, the present paper is the first to study how patient travel time (or distance) impacts \emph{both} the severity of the patient's health condition and the performance of a telemedicine system. Our results show that travel is indeed an important operational factor for hybrid health networks.
The stochastic model we use to capture the evolution of patients' health over time (in both hospitalization options and during travel) enables us to derive structural solutions and insights. These insights also address the question of whether and how telemedicine-based hospitalization can mitigate rural and non-rural healthcare disparities.}

\edit{Our work also contributes to the literature on health progression modeling, which has primarily focused on discrete state models. For instance,}
\cite{shi2021timing,deo2013improving}, and \cite{nambiar2020resource} explicitly modeled the individual patient progression by using a Markov chain model. \cite{grand2020robust} used an MDP to describe the evolution of patients' health condition and derive a proactive transfer policy to a hospital Intensive Care Unit (ICU). 
\cite{bavafa2019managing} modeled patient health dynamics using a Markovian continuous-time framework with three states: ``healthy,'' ``intermediate,'' and ``sick.'' \cite{bavafa2021customizing} analyzed primary care delivery through e-visits where patients become sick after an office visit, necessitating another visit after a random period, through a model with an increasing failure rate, linking longer intervals between visits to a higher sickness likelihood. More recently, \cite{bavafa2022surgical} introduced a model capturing patients' evolving health condition to study optimal discharge health, impacting readmission probability.
In this work, we also use a single aggregated health score to describe patients' health condition. Our model uses Brownian motion dynamics as the underlying mechanism to capture the dynamic evolution of health condition at each location. Being a BM model, it is fully characterized by its mean recovery speed (the drift) and variability (the diffusion coefficient), which can lead to deterioration. Modeling via drifted BMs has been used in sequential decision making and in the modeling of healthcare decisions (\citealp{siegmund2013sequential,wang2010design}), \edit{but, to the best of our knowledge, it has not yet been used to model the progression of the patient's health condition. Nevertheless, we will show how this model reproduces some LOS distributional assumptions commonly made in the literature, and this connection shows how our model parameters can be obtained from data}. We use the BM health score progression model to answer macro-level design questions, around which further refinement, such as dynamic control for individual patients, can be done.

\section{Modeling Hybrid Hospitalization and Patient Health Progression}
\label{sec: The model (new)}

Our modeling perspective in this work will operate on both micro- and macro-levels, capturing both the dynamics of each individual patient's health progression and the operational structure of the hybrid hospitalization network. Let us begin by describing the latter. 

\subsection{Two-Station Network of Remote and On-Site Hospitalization}

\begin{figure}[h]
	\caption{Illustration of the hybrid hospital service network stations.}
	\centering
	\includegraphics[width=0.7\textwidth]{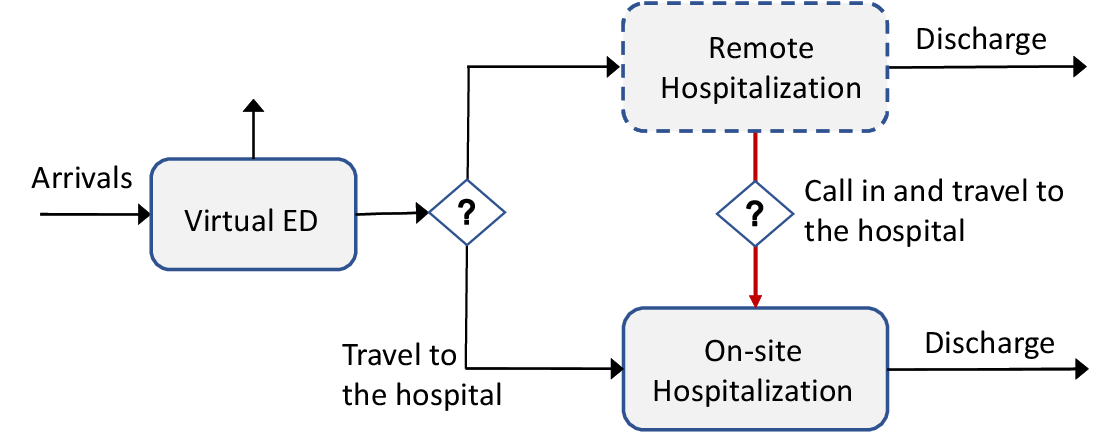}
	\label{fig: Model Illustration}
 \vspace{-.1in}
\end{figure}

Because the decision of whether to hospitalize a patient on-site or remotely must be made upon patient assessment, our hybrid hospital model begins after a triage through a virtual Emergency Department (ED). The full hospital network is depicted in Figure \ref{fig: Model Illustration}. After assessment at the virtual ED, patients can either be admitted remotely or advised to travel to the hospital for on-site admission. If admitted remotely, they either fully recover and are discharged, or, if their health condition worsens and reaches some predetermined threshold, they are called-in to the hospital and must travel to complete their hospitalization on-site.

We use the terms ``severity,'' ``health condition,'' or ``health score'' interchangeably in reference to a measure of {\em clinical acuity}.  The higher the score, the worse the health condition is.  Patients arrive to the virtual ED stochastically according to a renewal process with rate $\lambda$ and an \textit{initial health score} $x \in \mathbb{R}_+$. 
Upon arrival, a decision must be made as to whether to admit them remotely or on-site (following travel). If admitted on-site, they remain there until full recovery. If admitted remotely, they stay there as long as their health score does not reach a call-in threshold $x+a$, $a\geq0$. If a patient's health score reaches $0$ before it reaches $x+a$, they are discharged. Otherwise, when their score reaches $x+a$, they travel to the hospital, where they are admitted and stay there until they are healthy. 
We denote the travel time to the hospital by $T$. Note that if $a=0$, the patient is automatically admitted on-site, and if $a>0$, they are automatically first admitted remotely. Hence, the call-in threshold $a$ parsimoniously captures the health network's primary design decision: when should (or can) a patient be hospitalized remotely?

\subsection{Stochastic Dynamics of the Individual Health Score}
\label{subsec: Stochastic Dynamics of the Individual Health Score}

We model the evolution of patients health conditions through negative-drift BMs, which capture the recovery rates towards improvement during hospitalization as well as the randomness in recovery. Specifically, the health score of a patient during remote hospitalization is given by the process $$\calB^{R}(t)=x+\sigma_RB^{R}(t)-\theta_R t,$$ 
where $B^{R}(t)$ is a standard BM, $\theta_R > 0$ and $\sigma_{R}>0$. Thus, $\calB^{R}(t)$ is a negative-drift BM, starting at the initial score $x$, with drift $-\theta_{R}$ and diffusion coefficient $\sigma_{R}$. 

While the improvement rate at home being positive implies that home-hospitalized patients tend toward recovery and discharge, randomness allows the health score to increase, meaning that the patient's condition can become more severe. If a remotely hospitalized patient's condition deteriorates too much, they are called in to the hospital and complete the treatment there. Let $x+a$, $a>0$ denote the call-in threshold for a patient whose initial health score at admission was $x$. 
The remote hospitalization LOS is the first time a patient starting from health condition $x$ reaches health condition $0$ (discharge) or health condition $x+a$ (called-in) and is given by
$$\tau_{R}(x,a)=\inf\{t\geq 0:\calB^{R}(t)=0~ \text{or}~ \calB^{R}(t)=x+a \}.$$

The call-in likelihood, $\mathbb{P}\left(\calB^{R}(\tau_{R}(x,a))=x+a\right)$, is the probability that a patient who starts at health condition $x$ will reach health condition $x+a$ before reaching $0$. The expected LOS of remote hospitalization is $\Ex \left[\tau_{R}(x,a)   \right] $. Both have well known explicit expressions from the solution to the ``Gambler's ruin'' problem involving a BM. We have
\[
p_x(a) :=\mathbb{P}\left(\calB^{R}(\tau_{R}(x,a))=x+a\right)= \frac{1-e^{-\rho x}}{e^{\rho a} - e^{-\rho x}},
\]
where we define $\rho := 2\theta_R/\sigma_R^2 >0$, and 
\begin{align}\label{eq: remote LOS}
\Ex \left[\tau_{R}(x,a)   \right] = \frac{1}{\theta_R}
\left(\left(1-p_x(a)\right)x - p_x(a)a\right).
\end{align}


Patients who are called in have to travel to the hospital, and, naturally, their health condition may further degrade while traveling. We assume that their health score deteriorates according to a random variable, $Z(x,a,T)$ on $(-x-a,\infty)$, whose expected value is $T\theta_T$ for $\theta_T > 0$. Therefore, the patient's health score at arrival to the hospital is $x+a + Z(x,a,T)$. 


The model dynamics at the hospital are similar to the remote case, but with the difference of the initial starting health score being random, dependent on the patient's condition after the transit. The patient's health score's evolution is determined by 
$$\calB^{H}(t)=x+a+Z(x,a,T)+\sigma_H B^{H}(t)-\theta_H t,$$ 
where $B^{H}(t)$ is a standard BM, $\theta_H > 0$, and $\sigma_{H}>0$. We assume that the arrival process, $B^R,Z, \text{ and } B^H$ are independent.
Define
$$\tau_{H}(x,a,Z)=\inf\{t\geq 0:\calB^{H}(t)=0 \},$$
to be the patient's LOS at the hospital. Given $Z = Z(x,a,T)$, $\tau_{H}(x,a,Z)$ is the time it takes a BM with a negative drift $-\theta_H$, starting at $x+a+Z$ to hit zero. The expected LOS at the hospital is therefore
\begin{align}\label{eq: on-site LOS}
\Ex \left[\tau_{H}(x,a,Z)   \right] 
= 
\Ex \left[\Ex \left[\tau_{H}(x,a,Z) ~|~ Z\right]   \right] 
=
\frac{1}{\theta_H}
\Ex \left[\left( x+a+Z(a,x,T)\right)\right]=
\frac{1}{\theta_H}
\left( x+a+T \theta_T\right).
\end{align}

Finally, we complete the model by encoding a required clinical constraint, which enforces that the hospital never allows the patient to become too ill while being treated remotely. Let $\bar{S}$ be the most severe health condition allowed outside the hospital (in expectation). The call-in threshold then must satisfy that $x+a+T\theta_T \leq \bar{S}$. Letting $\bar{A}=  \left(0 \vee (\bar{S}-x-T\theta_T)\right)$, this policy constraint implies $a \in [0,\bar{A}] = \mathcal{A}$. 
Note that when $\bar{S}<x+T\theta_T$, the call-in threshold must be zero. 

\begin{remark}
\edit{Although we are interested in this BM model for its within-care representation of the patient's health progression, let us note that its hitting-time-based LOS actually reproduces the inverse Gaussian distribution already common in healthcare  modeling \citep[e.g.][]{whitmore1975inverse,hashimoto2023re}. In Appendix~\ref{sec: Parameter Estimation from Data}, we provide more details on this connection and use it to explain how the BM model parameters can be estimated from real-world healthcare data, motivating the data requirements of our model. \hfill \Halmos}
\end{remark}

\begin{figure}[h]
\caption{Three illustrative examples of patient's health score evolution. }
\centering
\includegraphics[width=\textwidth]{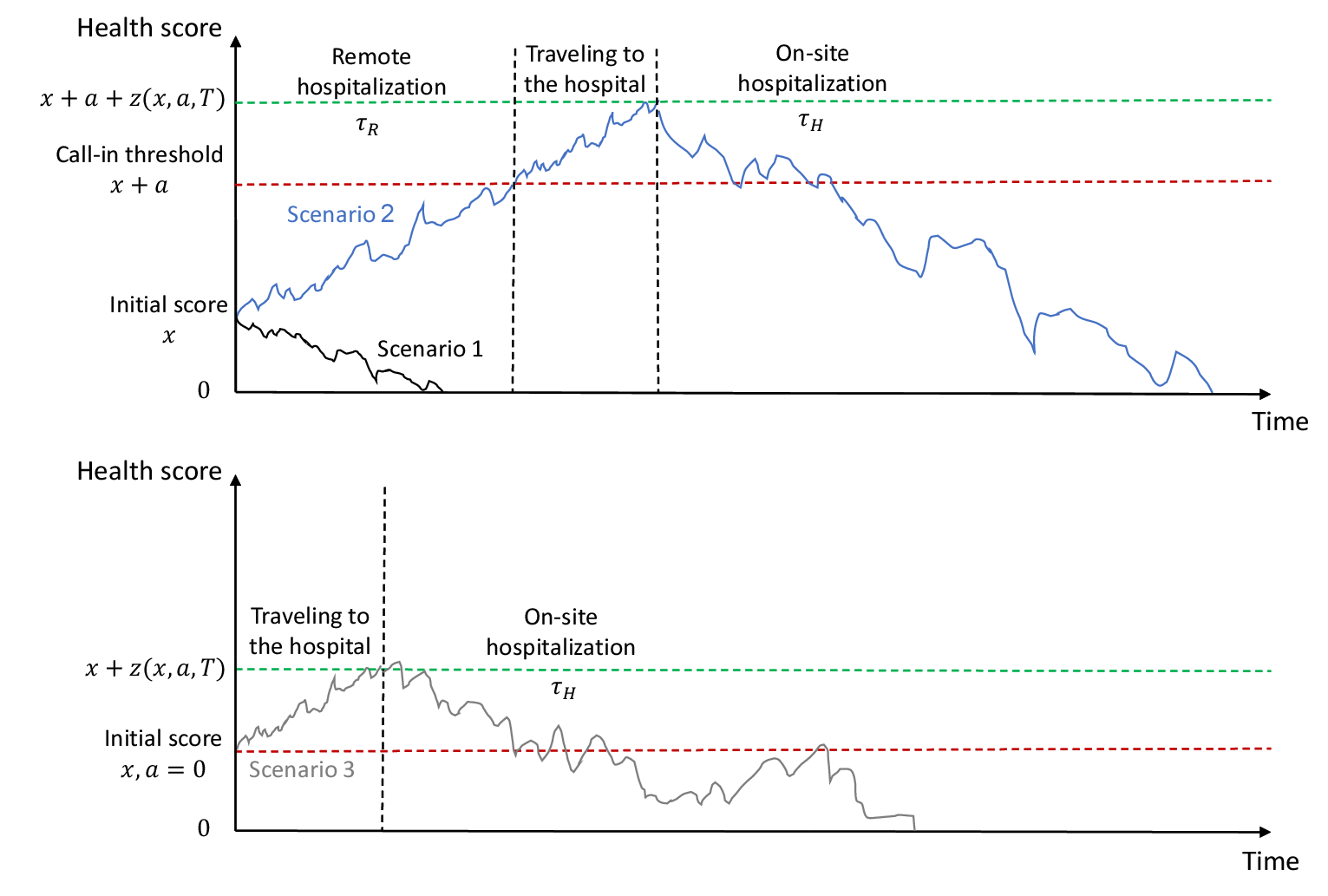}
\label{fig: Scenarios}
\vspace{-.45in}
\end{figure}

Figure \ref{fig: Scenarios} depicts three sample-path scenarios, all of which commence with a patient's health score at $x$ and involve a travel time of $T$ \edit{(if necessitated by the health progression)}. In Scenario 1, the patient is admitted remotely, improves\edit{,} and is discharged once their health score reaches zero.  
In Scenario 2, the patient is initially admitted remotely but experiences a decline in health. When the patient's health score reaches the predefined call-in threshold of $x+a$, they are called in to the hospital. During the journey to the hospital, the patient's health continues to deteriorate. Upon admission to the hospital, their health score is $x+a+Z(x,a,T)$, and from that point onward, the patient's condition improves.
In Scenario 3, the patient is called in to the hospital immediately upon arrival (i.e., $a=0$). Upon admission and travel to the hospital, the patient's health score is $x+Z(x,a,T)$, and from that point onward, the patient recovers on-site.

\subsection{Cost Structure and Optimization}
\label{subsec: Cost Structure and Optimization}

As mentioned, the health network's first-order design decision is captured in the threshold $x+a$. We expound upon that notion in this section. \edit{(In Section \ref{sec: Multiple Item Types}, we extend the results to accommodate multiple patient types that vie for limited resources.)} Because the negative drifts inherently capture health conditions that eventually improve, {the cost of care will be} the primary metric by which these decisions are assessed. Let $h_R$ and $h_H$ denote the holding cost rate for remote and on-site hospitalization, respectively. Similarly, $h_T$ denotes the cost rate \edit{during patient travel}. 
Accordingly, the total long run average cost is:
\begin{align}\label{eq:obj_func_linear}
V(a) 
&=  
\lambda\left(
h_R \Ex \left[\tau_{R}(x,a)   \right] 
+
\big(h_T T + h_H \Ex \left[\tau_{H}(x,a,Z)   \right] \big) p_x(a)
\right)
\nonumber
\\
&=
\lambda \left(\frac{h_R}{\theta_R}\left(\left(1-p_x(a)\right)x - p_x(a)a\right) + p_x(a)\left(h_T T + \frac{h_H}{\theta_H}\left(x+a+\theta_T T\right) \right)\right),
\end{align}
where our goal is to set the optimal call-in threshold $a \in \mathcal{A}$ that minimizes this cost. \edit{In Appendix \ref{subsec: Justification of Total Long-Run Average Cost}, we provide a mathematical justification for \eqref{eq:obj_func_linear} being the system's total long run average cost.} \edit{In Appendix \ref{sec: Incorporating Quadratic Holding Costs}, we consider the case where costs are quadratic, in which case the variance in recovery plays a more important role. However, in the main body of the paper, we restrict our attention to the objective in \eqref{eq:obj_func_linear}.}  

We find that it is useful to rewrite the value function (\ref{eq:obj_func_linear}) as 
\begin{align}\label{eq:obj_func_linear_equiv}
V(a)  =  \lambda\left(\alpha +\beta p_x(a) + \gamma p_x(a) a\right),
\end{align}
where the constants $\alpha$, $\beta$, and $\gamma$ are defined as follows:
\begin{align*}
\alpha &= h_Rx/\theta_R,\\
\beta &= -h_Rx/\theta_R + h_T T + h_H\left(x+\theta_T T\right)/\theta_H,\\
\gamma &= -h_R/\theta_R + h_H/\theta_H.
\end{align*}
Notice that $\beta = \gamma x + h_T T + h_H\theta_T T/\theta_H = \gamma x + (h_T + h_H\theta_T /\theta_H) T$.

In addition to the shorter expression, each of $\alpha$, $\beta$, and $\gamma$ offer interpretation to the decision problem. First, $\gamma$ represents the disparity in marginal costs between on-site and home hospitalization. Then, $\beta$ is the difference in expected costs of immediate transfer to on-site ($a=0$) and never transferring ($a=\infty$). Hence, $\beta$ measures the viability of immediate transfer versus exclusively doing remote hospitalization. Lastly, $\alpha$ is the expected cost of never transferring, or simply the expected cost per patient of exclusively doing home hospitalization: $V(0)/\lambda = \alpha + \beta$ and $V(\infty)/\lambda = \alpha$.

\vspace{3mm}

\textbf{Resource constraints.} The hospital has to allocate its resources, primarily medical staff, between two groups:  the on-site group which treats the on-site patients, and the virtual group, which is responsible for the remotely hospitalized patients. 
We start by defining the offered workload of each group.
The on-site workload is
\[
W_H(a) := \frac{\lambda p_x(a)}{\theta_H} \left(x+a+T\theta_T\right),
\]
whereas the remote workload is
\[
W_R(a) := \frac{\lambda}{\theta_R}\left(\left(1-p_x(a)\right)x - p_x(a)a\right). 
\]
The total workload is therefore
\[
W_{T}(a) := W_H(a)+W_R(a). 
\]

Consider a \textit{total} amount of resources, $C$, that needs to be allocated between the two groups. The corresponding optimization problem is
\be\label{eq:opt_RSharing}\begin{split}
\min_{a \in \mathcal{A} } V(a) = \min_{a \in \mathcal{A}}& \,\lambda\left(\alpha + \beta p_x(a) + \gamma p_x(a)a \right)\\
\mbox{s.t. } & 
W_{T}(a) \leq C.
\end{split} \ee
We denote the optimal call-in threshold by $a^*_{C}$ to emphasize the dependence of the solution on the total amount of resources. 
The solution (if it exists) to \eqref{eq:opt_RSharing} minimizes the cost $V(a)$, while balancing the on-site and remote workloads, $W_H$ and $W_R$, so that their sum does not exceed $C$. In particular, the dependence of $W_H$ and $W_R$ on $a$ dictates which thresholds allow the constraint in \eqref{eq:opt_RSharing} to be met and, therefore, encompasses the impact of resource scarcity. We elaborate on this in Section \ref{subsec:workload_analysis}.
In addition, the existence of a solution to \eqref{eq:opt_RSharing} depends on the problem parameters, and, in particular, the values of $\lambda$ and $C$. Indeed, if $\lambda$ is large and $C$ is small, the total workload constraint in \eqref{eq:opt_RSharing} might not be satisfied for any threshold $a \in \mathcal{A}$. Section \ref{subsec:Feasibility_Region} is devoted to characterizing the feasibility region in terms of $(\lambda,C)$ pairs. 

\begin{remark}
\edit{While our discussion and analysis will focus solely on the context of allocating finite resources, the constraint in~\eqref{eq:opt_RSharing} can also be motivated as restricting the approximate magnitude of waiting within the hybrid health network. For instance, suppose that \eqref{eq:opt_RSharing} is instead solved with the constraint $W_T(a) \leq \tilde C$ for some $\tilde C \leq C$, with $C$ as the true capacity. Then, the constraint from \eqref{eq:opt_RSharing} becomes equivalent to a bound on the \emph{inverse idleness:} $1/(C - W_T(a)) \leq 1/(C - \tilde C)$. The inverse idleness can be found in the \emph{utilization factor} of commonly used approximations for the wait within a $GI/G/C$ queue \citep[which, of course, is not available in closed-form; see, e.g.,][]{whitt1993approximations}. Although such moment-based approximations are known to be imprecise \citep{gupta2010inapproximability}, they are sufficient to demonstrate one of the most well-known principles in operations management: waiting grows exponentially as the utilization increases. Moreover, the way that these mean waiting formulas  capture this phenomenon is specifically through the utilization factor. Hence, by bounding $1/(C - W_T(a))$ through the choice of $\tilde C$, \eqref{eq:opt_RSharing} can be designed so that the solution minimizes the cost-of-care subject to limitations on the degree of the mean waiting. By similar reasoning around $\tilde C$ and $C$, \eqref{eq:opt_RSharing} can be designed so that the model reflects queueing theoretic properties of interest, such as aligning the utilization to that of a desired staffing regime. \hfill \Halmos}
\end{remark}

\section{Preliminary Analyses: Workload and Feasibility} 
\label{sec: Preliminary Results}

To identify the optimal call-in threshold and the resulting division of work among on-site and home hospitalization, we must first understand how the full operation depends on this level. In this pursuit, this section contains an analysis of the system's workload and a characterization of the problem's feasible region. 

\subsection{Analyzing the Shape of the Total Workload}\label{subsec:workload_analysis}
We begin by separately characterizing  the respective dependence of the on-site and remote hospitalization workloads on $a$. 

\begin{lemma}\label{lemma:W_on_a}
  $W_H(a)$ is a strictly decreasing function of $a$; $W_R(a)$ is a strictly increasing function of $a$. 
\end{lemma}

The intuition of Lemma \ref{lemma:W_on_a} is as follows: with an increase in the value of $a$, patients, on average, spend more time at home than in the hospital, by design. This translates into a rise in $W_R(a)$ and a decline in $W_H(a)$. The remaining question, tackled in Proposition \ref{prop:W_total_on_a}, pertains to the behavior of the sum $W_T(a) = W_H(a) + W_R(a)$ as a function of $a$. 
The pivotal factor influencing this behavior is the ratio of relative recovery rates: $\theta_H/\theta_R$.
Additionally, let $\Delta > 0$ be defined as
\begin{equation}\label{eq:def_of_Delta}
    \Delta
    =
    \frac{\rho \theta_T T} {  \rho x-1+e^{-\rho x}}
    .
\end{equation}
Through these two quantities, we can classify the shape of the workload as a function of the call-in threshold.

\begin{proposition}\label{prop:W_total_on_a}
The total workload $W_T(a)$ satisfies the following:    
\begin{enumerate}
\item \textbf{Case 1:} If $\theta_H/\theta_R\leq 1$, then $W_T(a)$ is strictly decreasing. 
\item \textbf{Case 2:} If $1<\theta_H/\theta_R<1+\Delta$, then $W_T(a)$ has a unique minimum $a_0$ in $(0,\infty)$. Moreover, $W_T(a)$ is strictly decreasing in $[0,a_0)$ and strictly increasing in $(a_0,\infty)$.
\item \textbf{Case 3:} If $\theta_H/\theta_R\geq 1+\Delta$, then $W_T(a)$ is strictly increasing. 
    \end{enumerate}
\end{proposition}

\begin{figure}[h!!]
\caption{An illustration of the total workload $W_T(a)$.}
\centering
\begin{subfigure}
\centering
\includegraphics[width=0.32\textwidth]{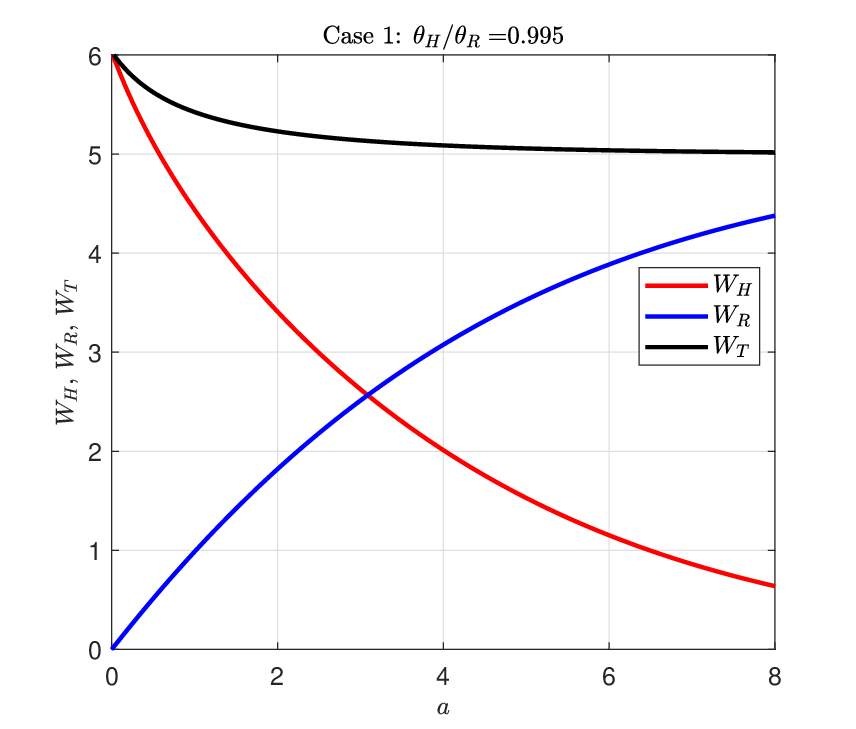}
\end{subfigure}
\centering
\begin{subfigure}
\centering
\includegraphics[width=0.32\textwidth]{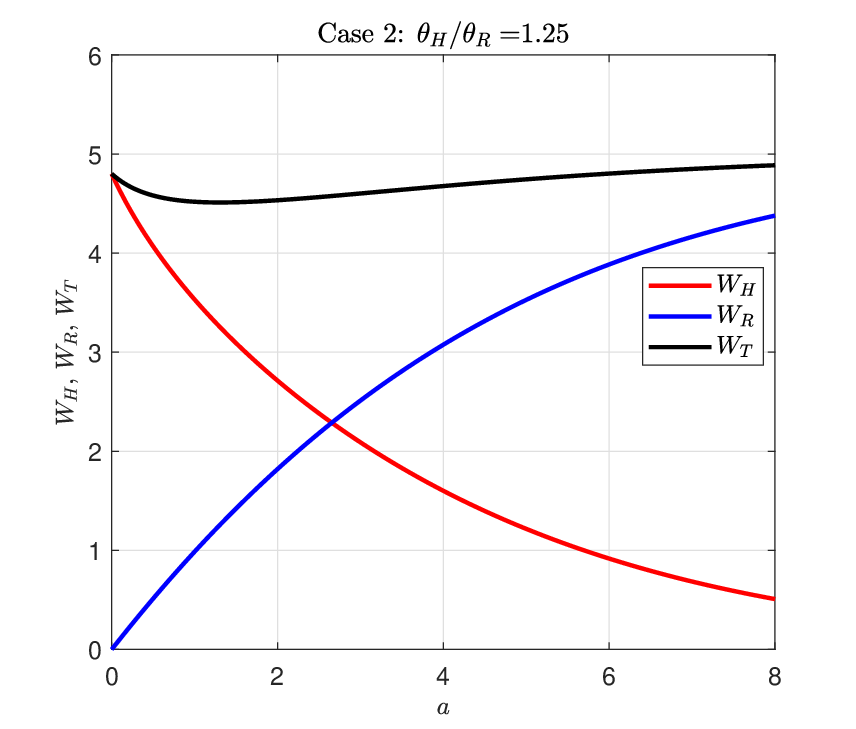}
\end{subfigure}
\centering
\begin{subfigure}
\centering
\includegraphics[width=0.32\textwidth]{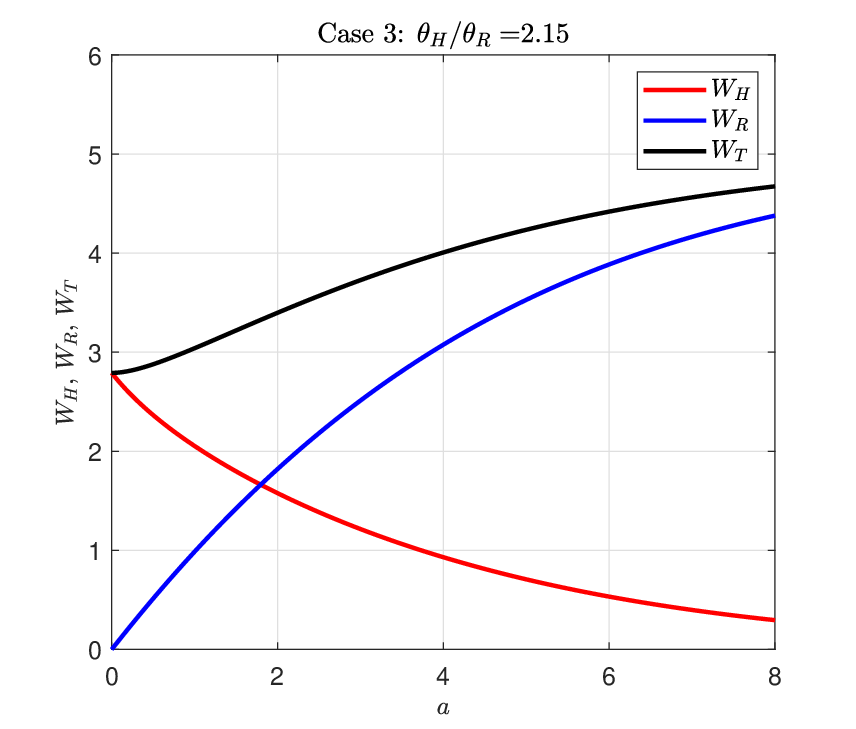}
\end{subfigure}
\label{fig: WT vs a}
\vspace{-.3in}
\end{figure}

From Proposition \ref{prop:W_total_on_a}, we see that the total workload $W_T(a)$ can have three forms. If the average recovery rate at the hospital is slower than at remote hospitalization (Case 1), minimizing the total workload can be achieved by increasing the call-in threshold to its maximum value. On the other hand, if the recovery rate at the hospital is much faster than under remote hospitalization (Case 3), minimizing the total workload is achieved by setting the call-in threshold to zero. Lastly, in the intermediate range when the on-site recovery rate is only moderately faster than under at home (Case 2), the total workload is unimodal with a unique minimum. Figure \ref{fig: WT vs a} illustrates these three cases.

Considering the problem context, Cases 2 and 3 each seem more realistic than Case 1, and Case 2 is likely the most interesting of all. First, it less likely that the average recovery rate at home would outpace  the full capabilities available at a hospital.  Then, under the same reasoning, it is of the greatest managerial intrigue to consider the setting when the hospital is indeed better on average, but only marginally so. 

The insights derived from Proposition~\ref{prop:W_total_on_a} will prove valuable in the upcoming section where we analyze the feasibility region. Furthermore, in Section \ref{subsec: The Capacitated Problem}, these insights will be instrumental as we analyze the capacitated solution. 


\subsection{Identifying the Feasibility Region}\label{subsec:Feasibility_Region}

Building on this understanding of the workload, let us now characterize, based on the given problem parameters, the $(\lambda, C)$ pairs for which there exists an $a \in \mathcal{A}$ satisfying the constraint in the optimization problem \eqref{eq:opt_RSharing}.
The feasibility region of \eqref{eq:opt_RSharing} is defined as: 
\[
{\calC_{FR}} = \left\{(\lambda,C) \in \mathbb{R}_+^2: \exists a \in \mathcal{A}, \text{ s.t. } W_T(a)\leq C \right\}.
\]
Let $a_{\min}$ denote the value of $a \in \mathcal{A}$ for which the total workload is minimal, i.e.,
\begin{align}
a_{\min}&=
\argmin_{a \in \mathcal{A}}W_T(a).
\label{aminDef}
\end{align}
Note that Proposition~\ref{prop:W_total_on_a} guarantees that $a_{\min}$ is unique. However, relative to the $a_0$ in Proposition~\ref{prop:W_total_on_a}, $a_{\min}$ is restricted to the range $\mathcal{A} = [0, \bar A]$, whereas $a_0 \in \mathbb{R}_+$.

Clearly, 
\edit{there exists} $a \in \mathcal{A}$  such that $W_T(a)\leq C$ \edit{if and only if the solution to \eqref{aminDef} is such that}  $W_T(a_{\min})\leq C$. Since $W_T(a_{\min})/\lambda$ does not depend on $\lambda$, and $a_{\min}$ minimizes it as well, we are essentially looking for $(\lambda,C)$ pairs such that $\lambda \left(W_T(a_{\min})/\lambda\right)\leq C$. 
Using this observation in tandem with Proposition \ref{prop:W_total_on_a}, we obtain the following characterization of the feasibility region.

\begin{proposition}\label{prop:feasibility_shared}
    The feasibility region of the optimization problem \eqref{eq:opt_RSharing} is given by:
\[
{\calC_{FR}} = \left\{(\lambda,C) \in \mathbb{R}_+^2:  W_T(a_{\min})\leq C \right\},
\]
where:
\begin{enumerate}
\item \textbf{Case 1:} If $\theta_H/\theta_R\leq 1$, then $a_{\min}=\bar{A}$. 
\item \textbf{Case 2:} If $1<\theta_H/\theta_R<1+\Delta$, then $a_{\min}=\min\{a_0 ,\bar{A}\}>0$, where $a_0$ is the unique minimum of $W_T(a)$ for $a \in \mathbb{R}_+$ (which does not depend on $\lambda$), as in Proposition \ref{prop:W_total_on_a}.
\item  \textbf{Case 3:} If $\theta_H/\theta_R\geq 1+\Delta$, then $a_{\min}=0$. 
    
\end{enumerate}
\end{proposition}

Proposition \ref{prop:feasibility_shared} characterizes the feasibility region by considering three cases, mirroring the distinctions established in Proposition \ref{prop:W_total_on_a} regarding the behavior of the total workload.
With this understanding of the structure of the arrival rates and capacities for which the resource allocation problem is feasible, let us now analyze the true optimization problem. 


\section{Minimizing the Cost-of-Care for Hybrid Hospitalization}
\label{sec: Results}

\edit{Following these preliminary analyses}, we are now prepared to address our focal  question of managing the operations of a hybrid hospital. To build insights, we will first study the unconstrained problem, where the resource capacity $C = \infty$, and we will then utilize this solution to analyze the finite $C$ case. \edit{Similarly, the insights we find for the single patient type setting in this section will guide our generalization to multiple types in the sequel.}

\subsection{Identifying the Optimal Call-In Structure with Unlimited Resources}
\label{subsec: The Uncapacitated Problem}

Recalling the simplified notation in Equation~\eqref{eq:obj_func_linear_equiv}, our present goal is to set the call-in threshold in order to minimize the total expected cost rate \edit{when only limited by the clinical boundary requirements on the patients health condition}:
\[
\min_{a \in \mathcal{A} } V(a) = \min_{a \in \mathcal{A}} \lambda\left[\alpha + \beta p_x(a) + \gamma p_x(a)a \right].
\]
Proposition \ref{prop:opt_a_New2} characterizes the uncapacitated optimal call-in threshold $a^*_{\infty}$. In particular, the extreme cases are $a^*_{\infty}=0$ and $a^*_{\infty}= \bar A = \bar{S}-T\theta_T-x$. When $a^*_{\infty}=0$, remote hospitalization is less cost effective than on-site hospitalization, and thus it is always preferable to hospitalize the patient on site. When $a^*_{\infty}=\bar A$, however, remote hospitalization is more cost effective, and thus it is always preferable to remotely hospitalize the patient until they reach the worst health condition that can still be treated remotely. \edit{Any $a_\infty^* \in (0, \bar A )$ lies between these extremes: the patient will start their hospitalization remotely, but the call-in threshold is lower than the maximum allowable.}

\begin{proposition}[optimal call-in threshold]\label{prop:opt_a_New2}
Let the travel time $T$ and initial condition $x>0$ be fixed. 
\begin{itemize}
\item If the marginal hospitalization cost is higher at the hospital ($\gamma\geq0$), then remote hospitalization is always preferable, and the call in threshold is as high as allowable ($a_{\infty}^*=\bar{A}$). 
\item If  the marginal hospitalization cost is smaller at the hospital ($\gamma<0$):
\begin{itemize}
    \item If immediate transfer to on-site is either not viable ($\beta \geq 0$) or viable but not dominant ($\gamma(1 - e^{-\rho x})/\rho < \beta < 0$), then the optimal threshold is given by $a_{\infty}^* = (\tilde a \wedge \bar{A})$, where $\tilde a > 0$ is the unique solution to
    \begin{align}
        e^{-\rho \tilde a}
        &=
        \left(
        1 - \beta \rho / \gamma - \rho \tilde a  
        \right)
        e^{\rho x}
        .
        \label{tildeAEq}
    \end{align}
    which can be expressed by
    \begin{align}
    \tilde a 
    &=
    \frac{1}{\rho}\left(
    1
    +
    \mathsf{W}\left(
    - e^{-\rho x + \beta \rho / \gamma - 1}
    \right)
    \right)
    -
    \frac{\beta}{\gamma}
    ,
    \label{solW}
    \end{align}
    with $\mathsf{W}(\cdot)$ as the Lambert-W function (principal branch).

    \item If immediate transfer is both viable and dominant ($\beta \leq \gamma(1 - e^{-\rho x})/\rho$), then $a_{\infty}^* = 0$; all patients are treated on-site and home hospitalization is not offered.
\end{itemize}
\end{itemize}
\end{proposition}

We find that $\gamma$ plays an important role in the decision of where to hospitalize patients and in case they were admitted, decide when to call them in to the hospital. Specifically, recall that $\gamma$ represents the marginal cost difference between on-site and remote hospitalization. If $\gamma$ is non-negative, then patients should be admitted to remote hospitalization and stay there as much as possible. \edit{If $\gamma$ is negative, on-site  hospitalization is more cost effective marginally, but remote hospitalization may still be preferable (at least initially) depending on the value of $\beta$. Recalling that $\beta$ measures the cost viability of immediate transfer, let us take a closer look at the impact of the patient's travel.}

\begin{figure}[h]
\caption{Optimal call-in threshold and call-in probability as a function of travel time for different initial health scores. The parameters are $\theta_H=0.05$, $\theta_R=0.06$, $\theta_T=0.1$, $h_H =2.65$, $h_R =5.1$, $h_T =2$, $\gamma = -32$, $\lambda = \sigma_R= 1$, $\bar{S}=15$.}
\centering
\begin{subfigure}
\centering
\includegraphics[width=0.491\textwidth]{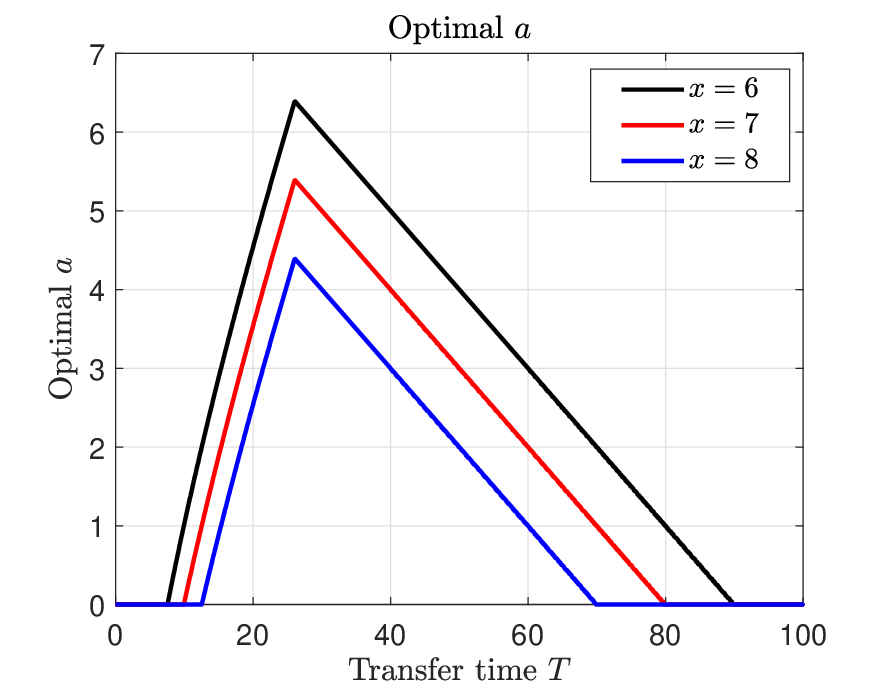}
\end{subfigure}
\centering
\begin{subfigure}
\centering
\includegraphics[width=0.491\textwidth]{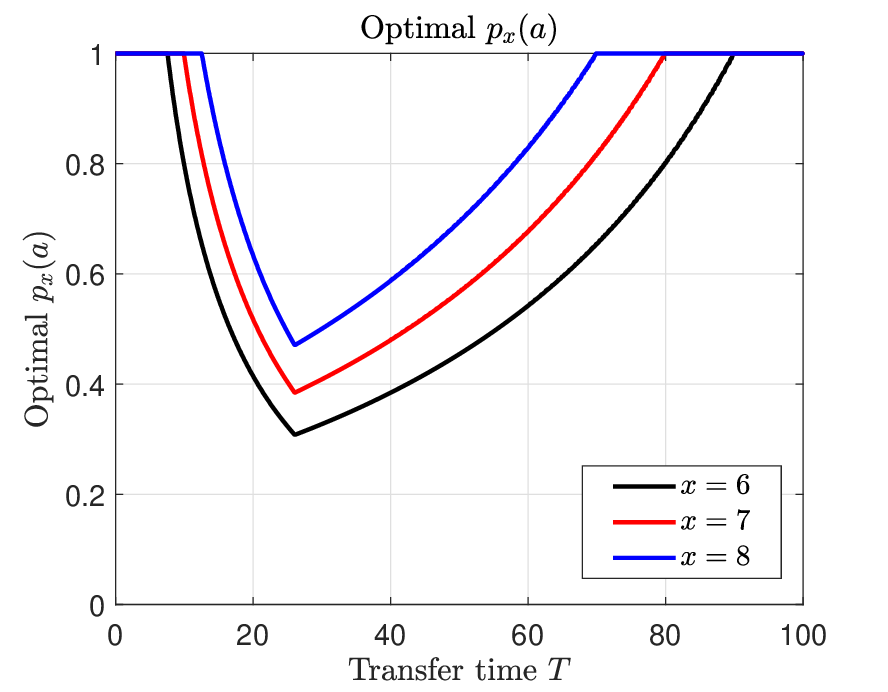}
\end{subfigure}
\label{fig: optimal a and p}
\vspace{-.4in}
\end{figure}

Figure \ref{fig: optimal a and p} illustrates the optimal call-in threshold and probability for different transfer times to the hospital and different initial conditions. First, we observe that remote hospitalization is not cost-effective for patients in close proximity or those residing at a significant distance from the hospital. In such cases, where $a_{\infty}=0$, direct on-site admission is deemed more appropriate. Moreover, the call-in threshold is not monotone in $T$: it initially rises, reaching a \edit{threshold-maximizing patient travel time}, $\hat{T}$ (which remains consistent across all initial severity levels), before subsequently declining back to zero. The call-in probability has the exact opposite structure: it starts at one, decreases\edit{,} and then increases back. Second, we can see that as the initial condition becomes more severe ($x$ increases), beyond the fact that the call-in threshold decreases, the distance range at which remote hospitalization is cost effective shrinks. Theorem~\ref{theorem: T properties} formalizes these properties. 

Specifically, to establish the decision's dependence on distance, let us first clarify how the model parameters depend on $T$. Recalling the definitions of $\alpha$, $\beta$, and $\gamma$ following Equation~\eqref{eq:obj_func_linear_equiv}, we can recognize that, among these, only $\beta$ depends on $T$. Moreover, if we define $\eta = h_T + h_H \theta_T / \theta_H$ as the marginal cost of travel distance, then $\beta$ can be simply re-expressed as $\beta = \gamma x + \eta T$. Exploiting this dependence, we formalize the observations from Figure~\ref{fig: optimal a and p} now in Theorem~\ref{theorem: T properties}.

\begin{theorem}
\label{theorem: T properties} 
Let $T_{LB}$ and $T_{UB}$ be defined such that
\begin{align}
T_{LB} &= -\frac{\gamma}{\eta}\left(x - \frac{1}{\rho}\left(1 - e^{-\rho x} \right)\right) 
\qquad
\text{ and }
\qquad
T_{UB} = \frac{\bar{S}-x}{\theta_T}. 
\end{align}
Then, \edit{if the marginal hospitalization cost is higher at the hospital ($\gamma\geq0$)}, $a_\infty^* > 0$ if and only if $T < T_{UB}$.

Furthermore, \edit{if the marginal hospitalization cost is higher at home ($\gamma<0$)}, then the $\hat T$ which is the unique solution to
\begin{align}
\bar{S}
&=
\frac{1}{\rho}
\left(
1
+
\mathsf{W}\left(
-e^{{\eta \rho }\hat T / \gamma - 1}
\right)
+
\left(\theta_T - \frac{\eta}{\gamma}\right)
\hat T
\right)
\label{ThatDef}
,
\end{align}
is such that for $T \in (T_{LB}, \hat T)$, 
\begin{align}
\frac{\partial a_\infty^*}{\partial T}
&=
-
\frac{\eta}
{\gamma}
\frac{
1
-
\mathsf{W}\left(
-
e^{\rho \eta T / \gamma - 1}
\right)
}
{
1
+
\mathsf{W}\left(
-
e^{\rho \eta T / \gamma - 1}
\right)
}
>
0
,\end{align}
and for $T \in (\hat T , T_{UB})$,
\begin{align}
\frac{\partial a_\infty^*}{\partial T}
&=
- \theta_T
<
0
,
\end{align}
with $a_\infty^* = 0$ for $T \not \in (T_{LB}, T_{UB})$.
\end{theorem}

Theorem~\ref{theorem: T properties} reveals that, even when remote hospitalization has lower marginal cost, the cost benefit only applies to patients up to a certain distance from the hospital. Though this may seem somewhat paradoxical at first glance, its intuition is clear: the maximum allowable patient severity ($\bar S$) and the distance-dependent expected deterioration while traveling ($\theta_T T > 0$) together imply that there is a ``shorter leash'' to risk on home hospitalization for patients who live far from the facility. This observation is further complicated by the recognition that the range of cases in which home hospitalization is viable narrows as the initial severity increases: $T_{LB}$ increases in $x$, $T_{UB}$ decreases, and the rates that $a_\infty^*$ changes with $T$ are exactly parallel across $x$ and thus are unaffected. Recalling the growing recognition of more dire health states in rural areas \citep{lewis2022people}, we see that Theorem~\ref{theorem: T properties} identifies a problematic combination. 
That is, if greater distance and worsened initial condition each restrict the feasibility of remote hospitalization, then this mode of care may not benefit the exact populations for which it seems intended. 

Let us emphasize that this conundrum is not a consequence of scarce resources -- thus far, our results have assumed an unlimited amount of resources. \edit{Of course, real-world health networks must manage hybrid hospitalization subject to resource limitations.} As we show in the next section, the capacitated solution retains the same properties to the uncapacitated counterpart. Consequently, the diminished effectiveness of remote hospitalization with distance and severity persists also in the presence of resource scarcity; in fact, in what may be the most realistic parameter settings, this reduction is exacerbated even further.


\subsection{Identifying the Optimal Call-In Structure with Limited Resources}
\label{subsec: The Capacitated Problem}
We now go back to our original capacitated problem in (\ref{eq:opt_RSharing}). The goal is two-fold. First, we wish set the call-in policy under finite amount of resources. 
Second, we wish to allocate the total amount of resources between the two hospitalization modes: on-site and remote.

To begin, Theorem \ref{theorem:opt_a_cap} characterizes the solution of the capacitated problem (\ref{eq:opt_RSharing}). 

\begin{theorem}\label{theorem:opt_a_cap}
    Assume that $W_T(a_{\min})\leq C$ \edit{for $a_{\min} \in \mathcal{A}$ as defined in \eqref{aminDef}} (i.e., the feasibility region is not empty).
    Then, problem \eqref{eq:opt_RSharing} has a unique solution $a_{C}^*\in \mathcal{A}$, such that:
    \begin{itemize}
        \item If $W_T(a_{\min})= C$, then $a_{C}^*=a_{\min}$.
        \item If $W_T(a_{\min})< C$, then:
        \begin{itemize}
            \item If $W_T(a_{\infty}^*)\leq C$, then $a_{C}^*=a_{\infty}^*$,
            \item If $W_T(a_{\infty}^*)> C$, then $a_{\min}\neq a_{\infty}^*$ and $a_{C}^*$ is the unique value of $a\in \mathcal{A}$ strictly between $a_{\min}$ and $a_{\infty}^*$ such that $W_T(a)=C$.
        \end{itemize}
    \end{itemize}
\end{theorem}

Note that depending on the parameters, both $a_{\min}>a_{\infty}^*$ and $a_{\min}<a_{\infty}^*$ are possible. In either case, when $W_T(a_{\infty}^*)> C$, and $W_T(a_{\min})< C$, the call-in threshold $a_{C}^*$ is strictly between them and satisfies (uniquely) $W_T(a_{C}^*)=C$. 

To interpret this structure and make its solution explicit, let us now establish an equivalence between the capacitated and uncapacitated solutions. To emphasize the dependence of the cost-of-care function $V$ on holding costs, we denote it as $V(h_R, h_H, a)$. Recall the uncapacitated minimization problem,
\be\label{eq:opt_uncap}\begin{split}
\min_{a \in \mathcal{A} } V(h_R,h_H,a),
\end{split} \ee
which per Proposition \ref{prop:opt_a_New2}, has a unique solution $a_{\infty}^*\in \mathcal{A}$. Recall also the capacitated minimization problem,
\be\label{eq:opt_RSharing_recall}\begin{split}
&\min_{a \in \mathcal{A} } V(h_R,h_H,a) \\
&\mbox{s.t. }  
W_{T}(a) \leq C,
\end{split} \ee
which per Theorem \ref{theorem:opt_a_cap}, assuming that $W_T(a_{\min})\leq C$, has a unique solution $a_{C}^*\in \mathcal{A}$.
Define
\begin{align}
\Gamma=
\begin{cases}
-\frac{V'(h_R,h_H,a_{C}^*)}{W_T'(a_{C}^*)}, ~~~\text{if }  W_T(a_{\min})< C \text{ and } W_T(a_{\infty}^*)>C\\
0, \hspace{23mm} ~\text{otherwise}.
\end{cases}
\label{GammaDef}
\end{align}
Note that $\Gamma\geq 0$, since in the case where $W_T(a_{\min})< C$ and $W_T(a_{\infty}^*)>C$, $V'(h_R,h_H,a_{C}^*)$ and $W_T'(a_{C}^*)$ must be non-zero and with opposite signs (see the proofs of Lemmas \ref{lemma:W_on_a} and \ref{lemma: prop V}). 
Now, consider a similar uncapacitated optimization problem with $\Gamma$-modified costs:
\be\label{eq:opt_RSharing_modified_costs}\begin{split}
\min_{a \in \mathcal{A} } V(h_R+\Gamma,h_H+\Gamma,a).
\end{split} \ee
Proposition \ref{prop:opt_Gamma} establishes the equivalence between the solutions of (\ref{eq:opt_RSharing_recall}) and (\ref{eq:opt_RSharing_modified_costs}). This equivalence implies that all properties of the uncapacitated problem apply to the capacitated problem. Notably, the solution structure, characterized by (modified) $\alpha, \beta$, and $\gamma$ as outlined in Proposition \ref{prop:opt_a_New2}, and the influence of patient travel distance on the optimal call-in policy, as indicated in Theorem~\ref{theorem: T properties}, remain consistent.

\begin{proposition}\label{prop:opt_Gamma}
    Assume that $W_T(a_{\min})< C$ (i.e. the feasibility region of (\ref{eq:opt_RSharing_recall}) contains more than one point).
    Then, the problem \eqref{eq:opt_RSharing_modified_costs} has a unique solution in $\mathcal{A}$ which equals $a_{C}^*$.
\end{proposition}

\begin{figure}[h]
\caption{\edit{Shifted call-in policy for different $\Gamma$ and $T$. The parameters are $\theta_T=0.1$, $h_R =5.1$, $h_T =2$, $x=8$, $\bar{S}=15$, $\lambda=\sigma_R=1$. In the left plot, $\theta_H=0.05$, $\theta_R=0.06$, $h_H =1$; on right, $\theta_H=0.1$, $\theta_R=0.05$, $h_H =7$.}}   
\centering
\begin{subfigure}
\centering
\includegraphics[width=0.491\textwidth]{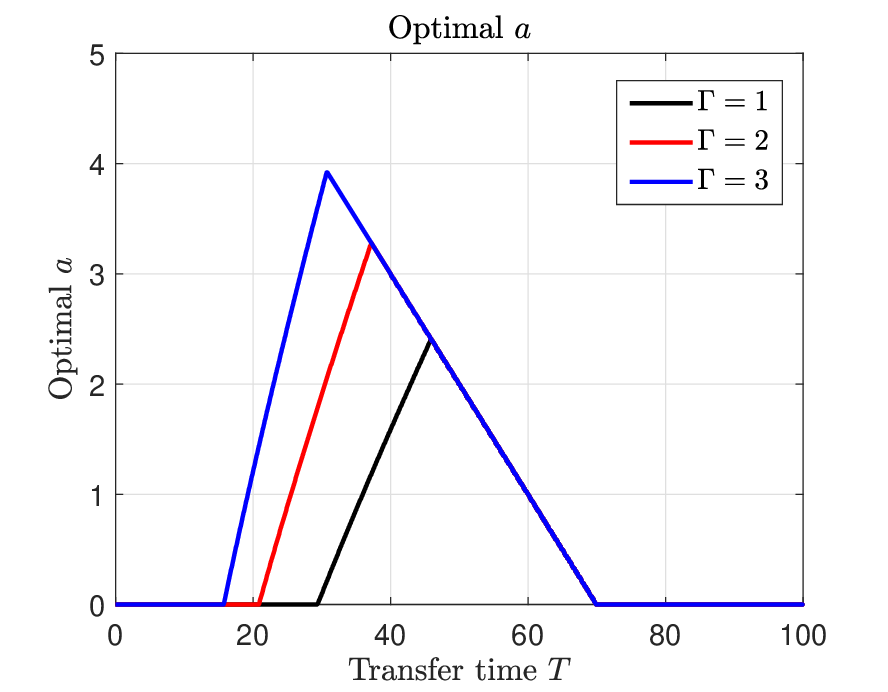}
\end{subfigure}
\begin{subfigure}
\centering
\includegraphics[width=0.491\textwidth]{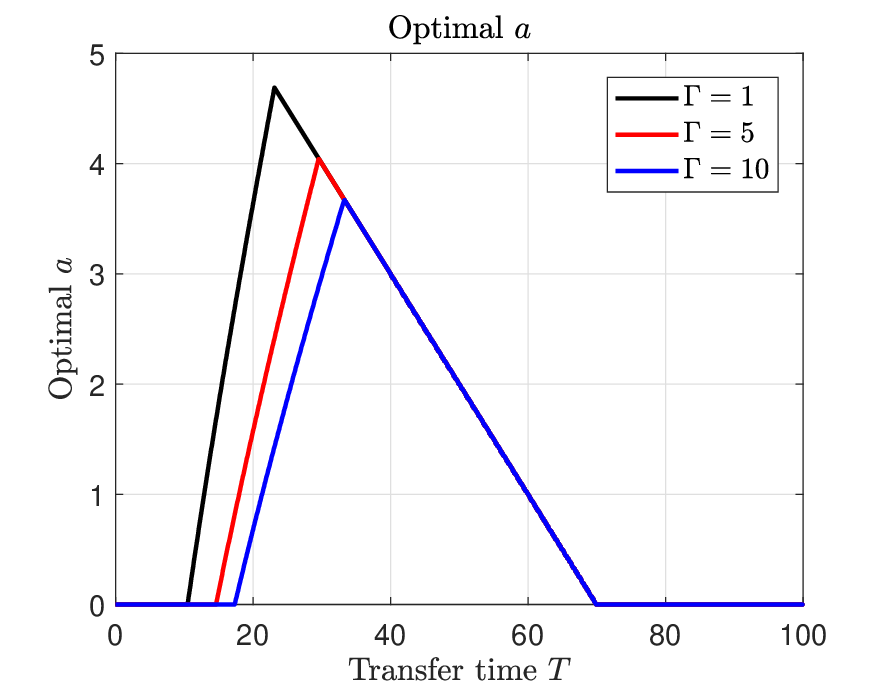}
\end{subfigure}
\label{fig: opt a - Diff Gamma and T}
\vspace{-.4in}
\end{figure}

The parameter $\Gamma$ captures the effect of scarce resources \edit{in a way that grants immediate managerial insights}. In essence, the capacity constraint effect is reflected through the simultaneous increase of both remote and on-site costs by $\Gamma$. Consequently, the call-in threshold will experience an increase or decrease contingent upon the initial cost rates and recovery rates associated with each hospitalization option. Specifically, the revised parameter $\gamma(\Gamma)$ would be 
\begin{align}
\gamma(\Gamma) = \frac{-(h_R+\Gamma)}{\theta_R} + \frac{h_H+\Gamma}{\theta_H} = \gamma + \Gamma\left(\frac{1}{\theta_H} - \frac{1}{\theta_R}   \right).
\label{gammaGammaEq}
\end{align}
Because $\Gamma\geq0$, the value of $\gamma(\Gamma)$ may increase {or} decrease depending on the relation between $\theta_H$ and $\theta_R$. Per Proposition \ref{prop:opt_a_New2}, the value of $\gamma(\Gamma)$, and in particular its sign, determines the optimal call-in threshold and whether, if at all, patients should be sent to remote hospitalization. 
When $\gamma>0$ and $\gamma(\Gamma) <0$, for example, patients who under ample resources would remain in remote hospitalization until their health score reaches \edit{$\bar{A}$}, would be called in at a lower threshold under a finite number of resources, or even directly admitted on-site. Furthermore, recalling Theorem~\ref{theorem: T properties}, we can notice that if $\theta_H > \theta_R$ and $\gamma < 0$, then the fact that $\gamma(\Gamma) < 0$ implies that an even smaller range of distances will be suitable for remote hospitalization, and this range again shrinks with the initial severity $x$. 
\edit{The right plot in Figure \ref{fig: opt a - Diff Gamma and T} illustrates this scenario. The figure presents the optimal call-in threshold as a function of $T$ for different values of $\Gamma$. (Note that by \eqref{GammaDef}, each $\Gamma$ corresponds to a different number of resources at each $T$.) Per Proposition \ref{prop:opt_Gamma}, the constrained solution is effectively equivalent to the unconstrained one with $\Gamma$-adjusted costs induced on both on-site and remote hospitalization (i.e., $h_R + \Gamma$ and $h_H + \Gamma$). In the right plot, as $\Gamma$ increases, the call-in threshold decreases, and the range at which remote hospitalization is cost-effective diminishes.}

\edit{Conversely, the left plot of Figure \ref{fig: opt a - Diff Gamma and T} demonstrates that as $\Gamma$ increases, the call-in threshold and the range where remote hospitalization is effective increase. Specifically, if $\theta_H < \theta_R$, then $\gamma(\Gamma) > \gamma$. When $\gamma < 0$ and $\gamma(\Gamma) > 0$, patients who, under ample resources, would be admitted on-site or remotely, would, under finite resources, be admitted remotely. For example, when $T = 25$ and $\Gamma = 1$, patients would be immediately called into the hospital, but when $\Gamma = 2,3$, they would first be admitted remotely.}





\begin{figure}[h]
\caption{Optimal capacitated solution. The parameters are $\theta_H=0.5$, $\theta_R=0.2$, $\theta_T=0.1$, $h_H =2.65$, $h_R =1.4$, $h_T =2$, $x=1$, $\bar{S}=15$, $\lambda=1$, $\sigma_R=1$.}
\centering
\begin{subfigure}
\centering
\includegraphics[width=0.46\textwidth]{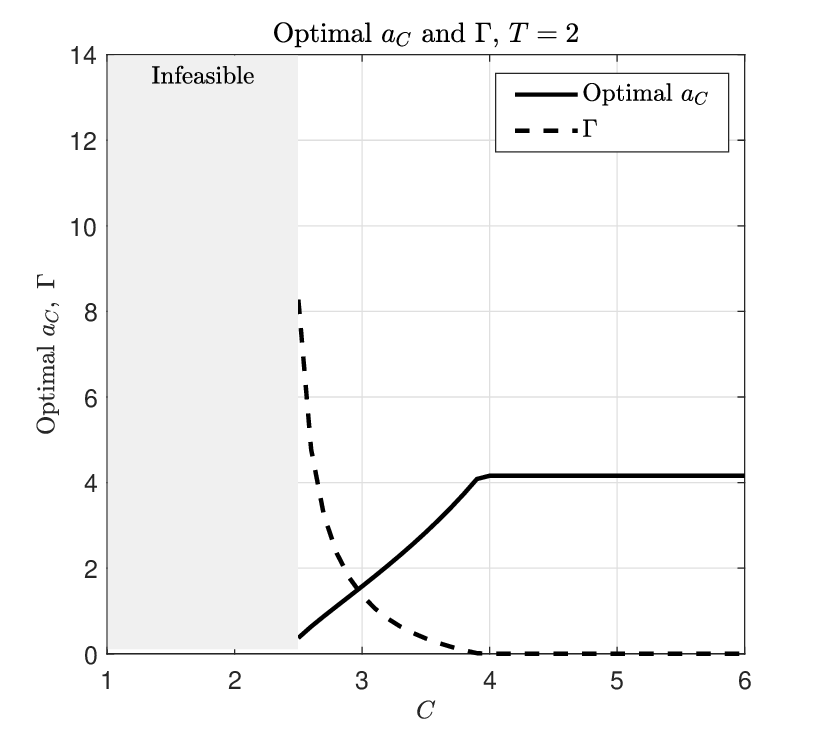}
\end{subfigure}
\begin{subfigure}
\centering
\includegraphics[width=0.46\textwidth]{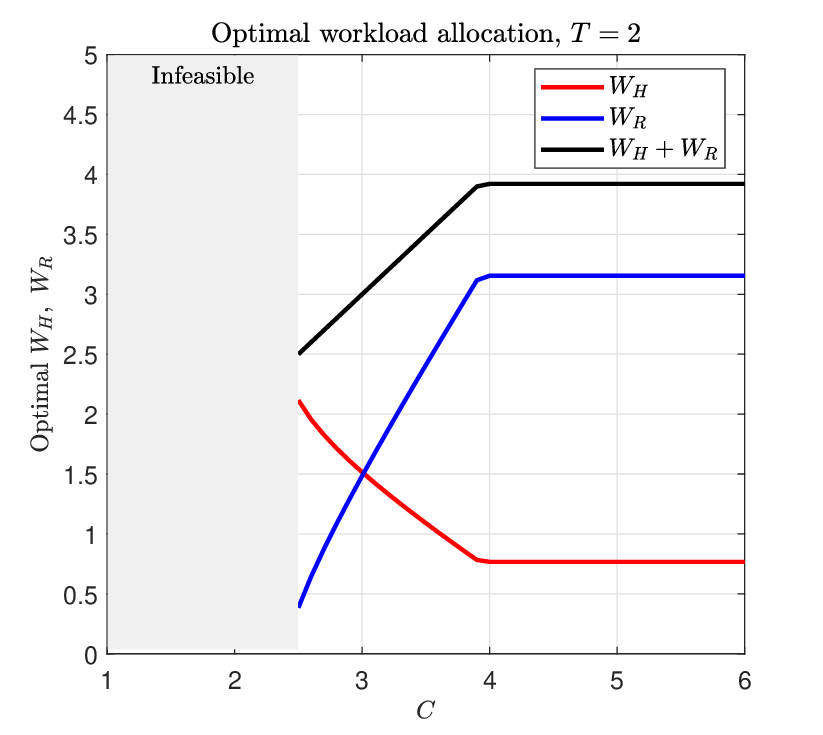}
\end{subfigure}
\centering
\begin{subfigure}
\centering
\includegraphics[width=0.46\textwidth]{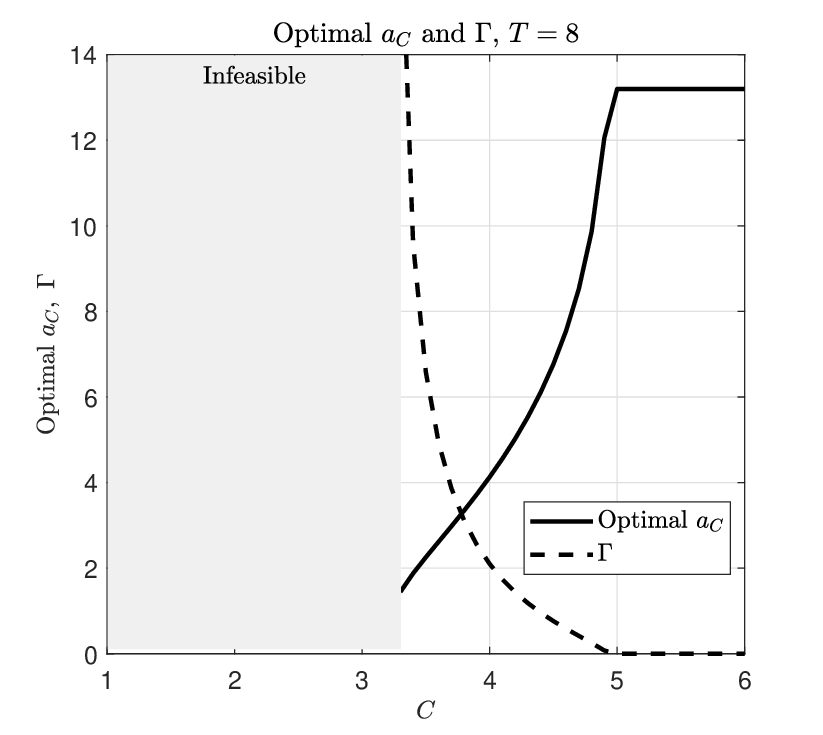}
\end{subfigure}
\begin{subfigure}
\centering
\includegraphics[width=0.46\textwidth]{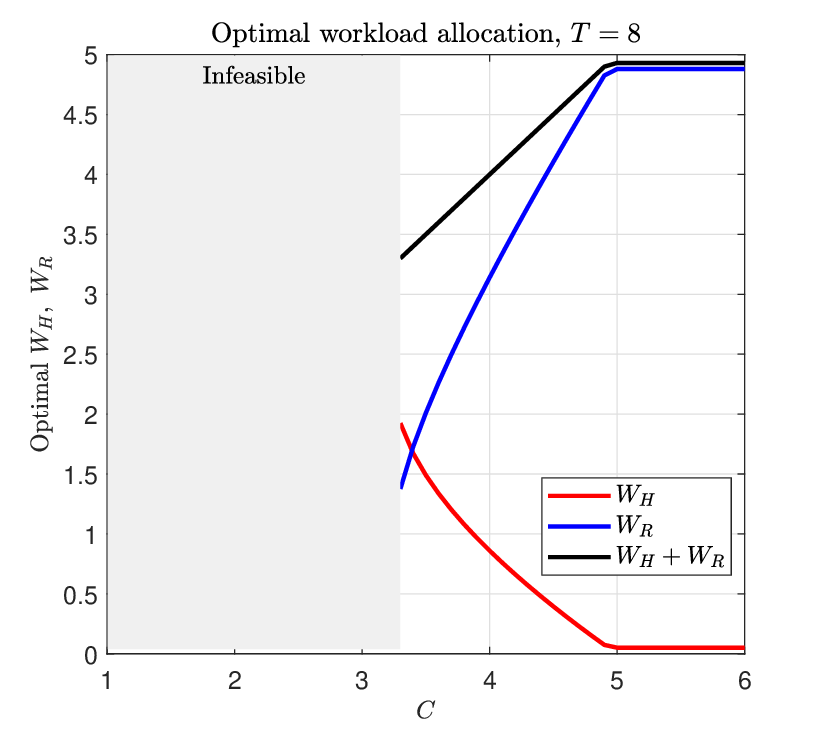}
\end{subfigure}
\label{fig: opt cap allocation - 1}
\vspace{-.25in}
\end{figure}

\begin{figure}[h]
\caption{\edit{Optimal capacitated solution. The parameters are $\theta_H=0.05$, $\theta_R=0.06$, $\theta_T=0.1$, $h_H =2.65$, $h_R =5.1$, $h_T =2$, $x=1$, $\bar{S}=10$, $\lambda=1$, $\sigma_R=1$.}}   
\centering
\begin{subfigure}
\centering
\includegraphics[width=0.46\textwidth]{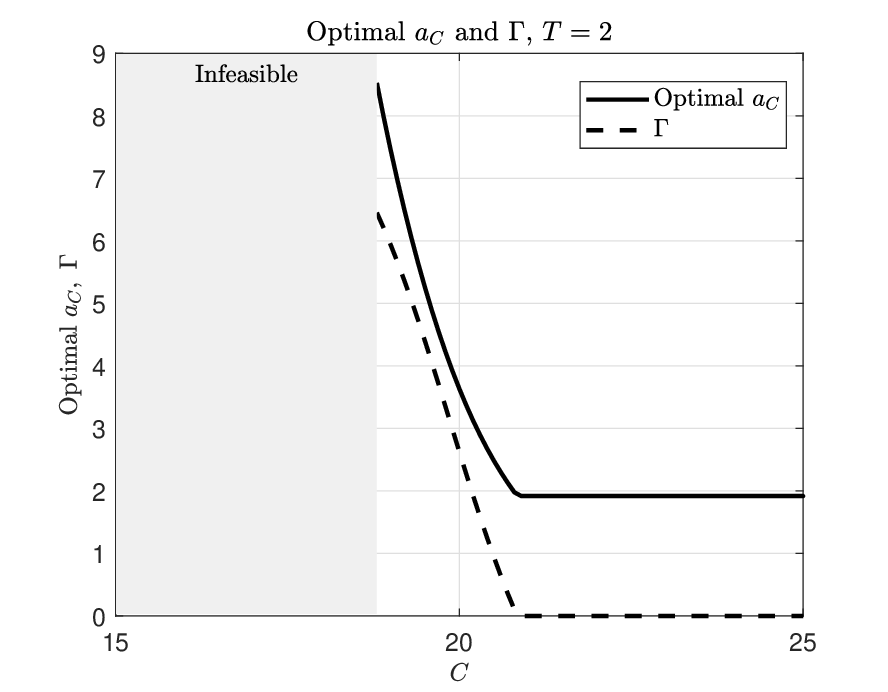}
\end{subfigure}
\begin{subfigure}
\centering
\includegraphics[width=0.46\textwidth]{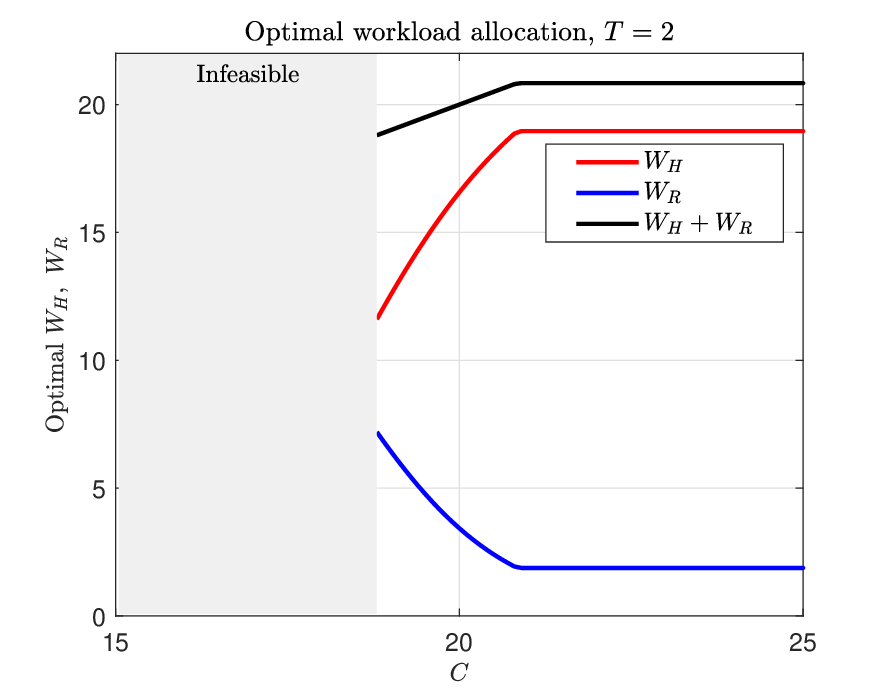}
\end{subfigure}
\label{fig: opt cap allocation - 2}
\vspace{-0.25in}
\end{figure}

Figures \ref{fig: opt cap allocation - 1} and \ref{fig: opt cap allocation - 2} illustrate the optimal capacitated solution for different resource levels and travel times. Figure \ref{fig: opt cap allocation - 1} corresponds to Case 3 in Proposition \ref{prop:W_total_on_a}; in the top plots ($T=2$): $2.5=\theta_H/\theta_R > \Delta+1 = 2.14$: When there is ample resources ($C\geq 4$), $a_{\infty}^*\approx 4$. As resources becomes scarce, the workload $W_T(a)$ decreases to satisfy the capacity constraint. Since in this case $W_T(a)$ is strictly increasing, the call-in threshold $a_C^*$ decreases up until $C\approx 2.5$, which is the boundary of the feasibility region. The bottom plots are for $T=8$. The feasibility region, which is smaller because $T$ is larger, ends at $C\approx 3.3$. In other words, more resources are needed when patients are distant. 

The right plots show the optimal resource allocation $W_H$ and $W_R$. We see that when resources are scarce, most of them ($80\%$ when $T=2$) are allocated to the hospital; as the total amount increases, fewer resources are allocated to the hospital, while more are allocated to remote hospitalization. When there are ample resources, most of them ($82\%$ when $T=2$) are allocated to remote hospitalization.

Figure \ref{fig: opt cap allocation - 2} corresponds to Case 1 in Proposition \ref{prop:W_total_on_a}, where $0.83 = \theta_H/\theta_R < 1$. In this case, $W_T(a)$ is strictly increasing. Therefore, when there is ample amount of resources, $a_{\infty}^*\approx 2$; as resources become scarce, $a_C^*$ increases. The right plot shows that, as opposed to the cases in Figure \ref{fig: opt cap allocation - 1}, as the total amount of resources increase, fewer are allocated to remote hospitalization, while more are allocated to the hospital.

\section{\edit{Generalizing the Insights to Multiple Patient Types}}
\label{sec: Multiple Item Types} 

\edit{
Thus far in the paper, we have obtained the optimal call-in threshold and resulting hybrid hospital workload allocation for a homogeneous stream of patients, building from the abundant resource setting to the constrained setting.
Of course, in practice, health networks must often manage and allocate medical resources across a heterogeneous patient population.
In this section, we generalize the analysis to incorporate multiple types of patients. When resources are scarce, the different patient types ``compete'' over them, implying that the hybrid hospital must optimally allocate \emph{across types}. As we will see, the analysis of the previous section provides a guidemap for this generalization, leading to a through-line of insights that spans from the single type setting to the multi-type.}

\edit{From here forward, we will now consider a population of patient types spanning indices $1, \dots, K$. We expand the prior notation by adding the superscript $k$ for type $k$ patients. Thus, for example, $x^k$ is the initial health score of type $k$ patients, $T^k$ is the travel time of type $k$ patients to the hospital, and $\theta_{H}^k$ is the rate of improvement at the hospital for type $k$. Let us also introduce  vector notation to denote the parameters for all types. For example, the  call-in vector for each patient type is denoted by $\vec{a} = a^1,\ldots,a^K$; $\vec{\mathcal{A}}$ stands for the $K$-dimensional set, where the $k$ component is the interval $\left[0,\bar{A}^k\right]$.} 

\edit{Like in~\eqref{eq:opt_RSharing}, let $C$ be the total amount of medical resources that must be allocated between the on-site and remote resource groups and, now, among the $K$ patient types.  The corresponding optimization problem is
\be\label{eq:opt_RSharing_Mult}\begin{split}
\min_{\vec{a} \in \vec{\mathcal{A}} } V(\vec{a}) = \min_{\vec{a} \in \bar{\mathcal{A}}}& \sum_{k=1}^K\lambda^k\left(\alpha^k + \beta^k p_x^k(a^k) + \gamma^k p_x^k(a^k) a^k \right)\\
\mbox{s.t. } & 
\sum_{k=1}^K W_{T}^k(a^k) \leq C.
\end{split} \ee
The vector of optimal call-in threshold is, therefore, $\vec{a}^*_{C}$. 
Like in the single type setting, the solution (if it exists) to \eqref{eq:opt_RSharing_Mult} minimizes the cost $V(\vec{a})$, while balancing the total on-site and remote workloads, $W^k_H$ and $W^k_R$, for each patient type $k=1,\ldots,K$, so that their total sum does not exceed $C$. In particular, the dependence of $W_H^k$ and $W_R^k$ on $a^k$, established in Section \ref{subsec:workload_analysis}, determines for which thresholds the constraint in \eqref{eq:opt_RSharing_Mult} is satisfied. 
The existence of a solution to \eqref{eq:opt_RSharing_Mult} depends on the problem parameters, and, in particular, the values of $\vec{\lambda}$ and $C$. Indeed, if the components of $\vec{\lambda}$ are large and $C$ is small, the total workload constraint in \eqref{eq:opt_RSharing_Mult} might not be satisfied for any threshold $\vec{a} \in \vec{\mathcal{A}}$. }


\subsection{\edit{Characterizing the Feasibility Region in Multiple Patient Dimensions}}
\label{subsec: The Multi-Type Feasibility Region}

\edit{Based on the given problem parameters, we characterize the feasibility region, meaning the $(\vec{\lambda}, C)$ for which there exists an $\vec{a} \in \vec{\mathcal{A}}$ satisfying the constraint in the optimization problem \eqref{eq:opt_RSharing_Mult}.
Formally, this is defined as: 
\[
{\calC_{FR}^K} = \left\{(\vec{\lambda},C) \in \mathbb{R}_+^{K+1}:  \exists 
\vec{a}\in \vec{\mathcal{A}}\text{ s.t. }\sum_{k=1}^K W_T^k(a^k)\leq C \right\}.
\]
We can now make a series of observations for the multi-type workload that essentially mirror what we saw for the single-type feasibility region in Section~\ref{subsec:Feasibility_Region}.
Let $a_{\min}^k$, $k=1,\ldots,K$, denote the value of $a^k \in \mathcal{A}^k$ for which the total workload is minimal, i.e.,
$$a_{\min}^k=\argmin_{a^k \in \mathcal{A}^k}W_T^k(a^k).$$
Note that Proposition~\ref{prop:W_total_on_a} guarantees that $a_{\min}^k$ is unique. However, relative to the $a_0^k$ in Proposition~\ref{prop:W_total_on_a}, each $a_{\min}^k$ is restricted to the range $\mathcal{A}^k = [0, \bar{A}^k]$, whereas $a_0^k \in \mathbb{R}_+$. 
By definition, there exists
$\vec{a} \in \vec{\mathcal{A}}$ 
such that $\sum_{k=1}^K W_T^k(a^k) \leq C$ if and only if  $\sum_{k=1}^K W_T^k(a_{\min}^k)\leq C$. Since $W_T^k(a_{\min}^k)/\lambda^k$ does not depend on $\lambda^k$, and $a_{\min}^k$ minimizes it as well, we are essentially looking for $(\vec{\lambda},C)$ such that $\sum_{k=1}^K\lambda^k \left(W_T^k(a_{\min}^k)/\lambda^k\right)\leq C$.}

\edit{Using this sequence of observations and Proposition \ref{prop:W_total_on_a}, we obtain the following characterization of the multi-type feasibility region. Like we first saw in Proposition~\ref{prop:feasibility_shared}, the structure of the minimal workload for each type will depend on the comparison of the type-specific recovery rates, $\theta_H^k$ and $\theta_R^k$, and the type-specific distance-dependent threshold, $\Delta^k = \rho^k \theta_T^k T^k/(\rho^k x^k - 1 + \exp(-\rho^k x^k))$, as generalized from \eqref{eq:def_of_Delta}.}

\begin{proposition}\label{prop:feasibility_shared_Mult}
\edit{The feasibility region of the optimization problem \eqref{eq:opt_RSharing_Mult} is given by:
\[
{\calC_{FR}^K} = \left\{(\vec{\lambda},C) \in \mathbb{R}_+^{K+1}:  \sum_{k=1}^K W_T^k(a_{\min}^k)\leq C\right\},
\]
where for each $k=1,\ldots,K$:}
\begin{enumerate}
\item \edit{\textbf{Case 1:} If $\theta_H^k/\theta_R^k\leq 1$, then $a_{\min}^k=\bar{A}^k$.} 
\item \edit{\textbf{Case 2:} If $1<\theta_H^k/\theta_R^k<1+\Delta^k$, then $a_{\min}^k=\min\{a_0^k ,\bar{A}^k\}>0$, where $a_0^k$ is the unique minimum of $W_T^k(a^k)$ for $a^k \in \mathbb{R}_+$ (which does not depend on $\lambda^k$), as in Proposition \ref{prop:W_total_on_a}.}
\item  \edit{\textbf{Case 3:} If $\theta_H^k/\theta_R^k\geq 1+\Delta^k$, then $a_{\min}^k=0$.}     
\end{enumerate}
\end{proposition}

\edit{Much like Proposition~\ref{prop:feasibility_shared}, the structural benefit of Proposition~\ref{prop:feasibility_shared_Mult} is that understanding the multi-type feasibility region has been reduced to understanding the workload minimizing call-in levels, which themselves can be characterized by the relationship between $\theta_H^k$, $\theta_R^k$, and $\Delta^k$. This prepares us to identify the structure of the optimal call-in thresholds for the generalized multiple-type, resource-limited setting.}

\subsection{\edit{Scaffolding to the Optimal Multi-Type Solution}}
\label{subsec: The Optimal Multi-Type Solution}
\edit{Throughout this section, we consider cases where the feasibility region contains more than one point, implying it is infinite. Given the non-convex nature of \eqref{eq:opt_RSharing_Mult}, both in its objective function and capacity constraint, finding a global optimal solution is challenging. However, by leveraging the problem's structure, we can gain valuable insights into its solution. We begin by proving the existence of a global optimal solution.}

\begin{lemma}\label{lemma: multi type existence}
    \edit{If the feasibility region is not empty, then there exists a feasible vector $\vec{a}^*_C$ which is the solution to the optimization problem \eqref{eq:opt_RSharing_Mult}.}
\end{lemma}

\edit{Next, we differentiate between two cases: when the vector of optimal call-in thresholds with unlimited resources is within the feasibility region and when it is not. For the first case, we have the following result.}
 \begin{lemma}\label{lemma: multi type optimality if in FR}
    \edit{If $\vec{a}^*_\infty \in \cal{C}_{FR}^K$, then $\vec{a}^*_\infty $ is the unique solution of the optimization problem.}
\end{lemma}

\edit{Lemma~\ref{lemma: multi type optimality if in FR} states the intuitive and desirable fact that if we can select the optimal solution for each patient type while still satisfying the resource constraint, then this would be the optimal outcome. However, if this threshold choice is not feasible, we must adjust the thresholds for some or all types, reducing their total workloads until the capacity constraint is met. Much like how Theorem~\ref{theorem:opt_a_cap} and Proposition~\ref{prop:opt_Gamma} identified the structure of the optimal call-in threshold for the single-type setting, Theorem \ref{theorem: multi type sol char} characterizes the solution for the second case, in which the abundant resource solution is not attainable.}

\begin{theorem}\label{theorem: multi type sol char}
    \edit{Suppose $\vec{a}^*_\infty \notin \cal{C}_{FR}^K$ and let $\vec{a}^*_C$ be a solution to the optimization problem (\ref{eq:opt_RSharing_Mult}). Then:}
    \begin{enumerate}
    \item \edit{Each threshold in $\vec{a}^*_C$ must lie between (and can be equal to) $a_{\min}^k$ and $a_{\infty,k}^*$ of the corresponding type.}
    \item \edit{The resource constraint at $\vec{a}^*_C$ is active, i.e., the sum of the total workloads at $\vec{a}^*_C$ is exactly $C$.}
    \item \edit{Denote by $E$ the set of all indices for which $[\vec{a}^*_C]_k \notin \{a_{\min}^k,a_{\infty,k}^*\}$. Then:}
    \begin{itemize}
        \item \edit{There exist a unique $\Gamma>0$ such that $\Gamma =-{V'}_k([\vec{a}^*_C]_k)/{W^k_T}'([\vec{a}^*_C]_k)$, for all $k\in E$.} 
        \item \edit{Specifically, $\vec{a}^*_C$ restricted to the entries in $E$ is the unique solution to the unconstrained optimization problem: $\min_{\vec{a} \in \vec{\mathcal{A}}_{E} } \sum_{k\in E}V_k\left(h_R+\Gamma,h_H+\Gamma,a_k\right)$, where $\vec{\mathcal{A}}_{E}$ is the boundary restricted to the entries in $E$.}
    \end{itemize}
    \end{enumerate}
\end{theorem}

\edit{Theorem \ref{theorem: multi type sol char} establishes that if the optimal solution for a patient type does not lie on the boundaries $a_{\min}^k$ or $a_{\infty,k}^*$, it is equivalent to the unconstrained solution, but with costs modified by a factor of $\Gamma$, which is consistent across all such types. Like in the single type setting,  the parameter $\Gamma$ captures the impact of resource scarcity on different patient types as an increase in both remote and on-site costs by $\Gamma$. This adjustment results in the call-in threshold either increasing or decreasing, depending on the initial cost rates and recovery rates specific to each type and hospitalization option. In the following section, we present compelling numerical examples to illustrate this effect.}


\subsection{\edit{Numerical Examples for the Multi-Type Problem}}
\label{subsec: Numerical Examples for the Multi-Type Solution}

\begin{figure}[h]
\caption{\edit{Optimal capacitated solution with two patient types. In the top plots: $(\theta_H^1,\theta_H^2)=(0.35,0.5)$, $(\theta_R^1,\theta_R^2)=(0.25,0.2)$, $(\theta_T^1,\theta_T^2)=(0.1,0.1)$, $(h_H^1,h_H^2) =(2.65,3)$, $(h_R^1,h_R^2) =(1.4,1.4)$, $(h_T^1,h_T^2)=(2,2)$, $(x^1,x^2)=(1,1)$, $(\bar{S}^1,\bar{S}^2)=(12,15)$, $(\lambda^1,\lambda^2)=(1,1)$, $(\sigma_R^1,\sigma_R^2)=(1,1)$, $(T^1,T^2)=(8,5)$. In the bottom plots: $(\theta_H^1,\theta_H^2)=(0.5,0.05)$, $(\theta_R^1,\theta_R^2)=(0.2,0.06)$, $(\theta_T^1,\theta_T^2)=(0.1,0.1)$, $(h_H^1,h_H^2) = (2.65,2.65)$, $(h_R^1,h_R^2) =(1.4,5.1)$, $(h_T^1,h_T^2) =(2,2)$, $(x^1,x^2)=(1,1)$, $(\bar{S}^1,\bar{S}^2)=(15,15)$, $(\lambda^1,\lambda^2)=(1,1)$, $(\sigma_R^1,\sigma_R^2)=(1,1)$, $(T^1,T^2) = (2,2)$.}}    
\centering
\begin{subfigure}
\centering
\includegraphics[width=0.491\textwidth]{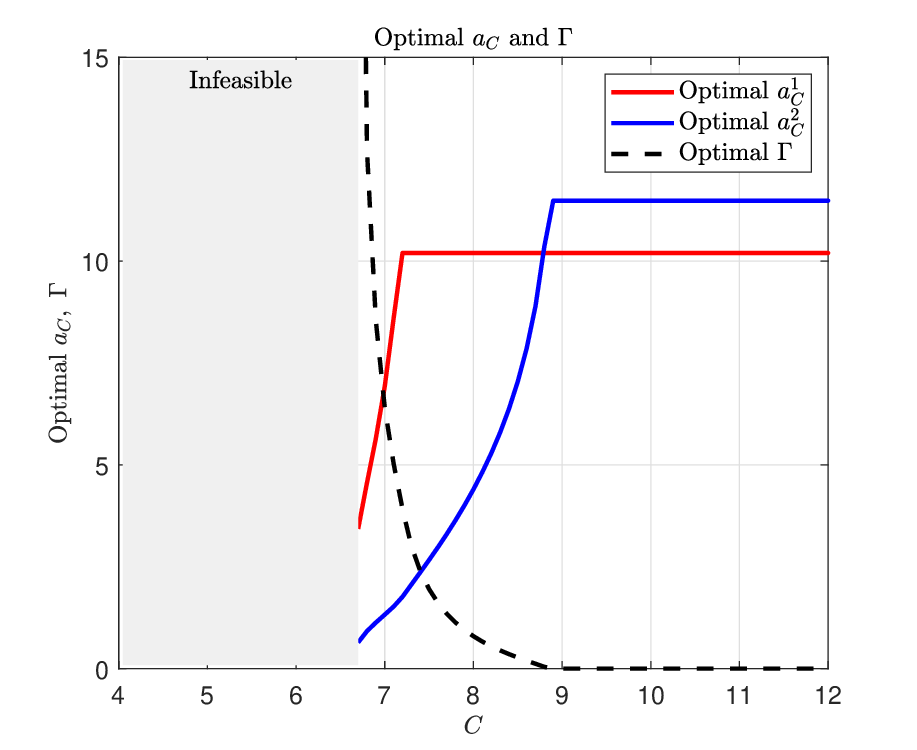}
\end{subfigure}
\begin{subfigure}
\centering
\includegraphics[width=0.491\textwidth]{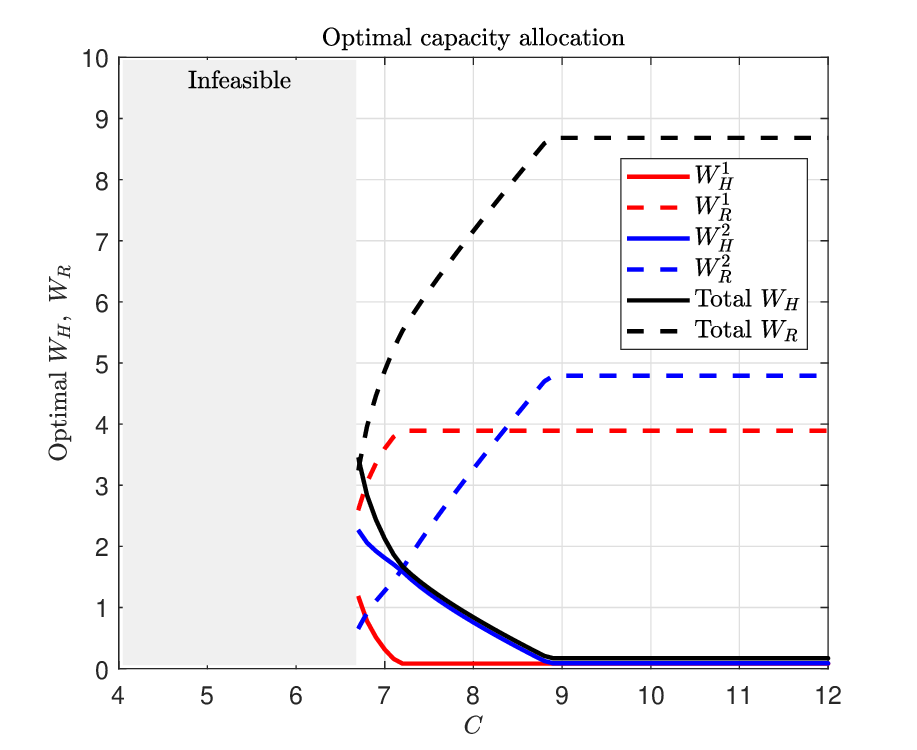}
\end{subfigure}
\centering
\begin{subfigure}
\centering
\includegraphics[width=0.491\textwidth]{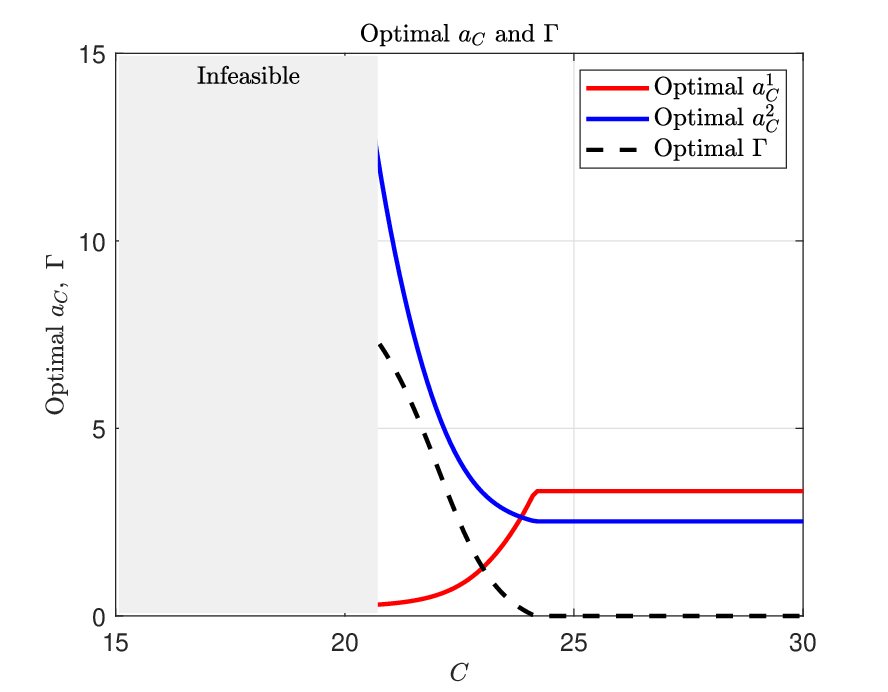}
\end{subfigure}
\begin{subfigure}
\centering
\includegraphics[width=0.491\textwidth]{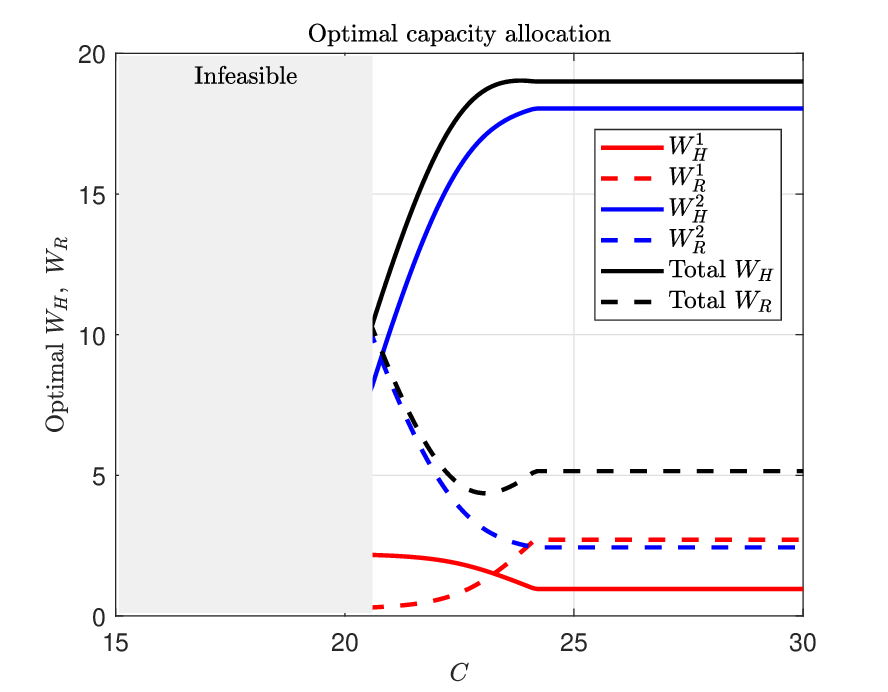}
\end{subfigure}
\label{fig: opt cap allocation - Mult}
\vspace{-0.3in}
\end{figure}

\edit{Let us now provide two examples that illustrate the optimal multi-type solution — call-back thresholds and resource allocation — for two patient types. In the first example illustrated in the top plots of Figure \ref{fig: opt cap allocation - Mult}, the two patient types differ, among other parameters, in their distance from the hospital. Type 1 is the distant one while Type 2 is closer. As could have been anticipated from the single-type analysis in Section \ref{subsec: The Uncapacitated Problem}, when there are ample resources, Type 1 has a smaller call-in threshold than Type 2. However, as resources becomes scarce, it is no longer possible to provide the optimal treatment mix to all. Since both types fall into Case 3 in Proposition \ref{prop:W_total_on_a} (i.e., $W_T(a)$ is strictly increasing), the call-in thresholds start decreasing. Interestingly, when $C\approx8.8$, the thresholds switch, so that Type 2 (the distant one) now has a larger call-in threshold. In other words, the distant patient now stays longer in remote hospitalization, while the closer one is called back earlier in a better health condition. Since both thresholds decrease as resources becomes scarce, the on-site workload increases, whereas the remote workload decreases.}

\edit{In the second example, illustrated in the bottom plots of Figure \ref{fig: opt cap allocation - Mult}, we consider two other patient types.  Interestingly, as resources become scarce, the threshold of Type 1 decreases (i.e., these patients are called into the hospital at a better health condition), while the threshold of Type 2 increases (i.e., these patients stay longer at home and are called into the hospital at a worse health condition). The reason for this phenomenon is that each type falls into a different case in Proposition \ref{prop:W_total_on_a}: Class 1 falls into Case 1 (i.e., $W_T(a)$ is strictly decreasing) and Class 2 falls into Case 3 (i.e., $W_T(a)$ is strictly increasing). Practically, this means that the call-in policy changes with the availability of resources, such that each type's call-in threshold may increase or decrease, depending on the case in Proposition \ref{prop:W_total_on_a}
 to which they belong. Moreover, the total workload allocation to on-site and remote hospitalization may be non-monotonic, as the allocation for each patient type moves in opposite directions.}


\section{\edit{Demonstration in Online Decision Making: A Dynamic Swap Policy}}
\label{subsec: A Dynamic Policy}

\edit{Thus far, we have focused on understanding the first-order design questions of hybrid hospitals; specifically, setting the call-in thresholds for each patient type and allocating resources between on-site and remote hospitalization. After setting the optimal system design, the second-order questions include dynamic control policies that support decision making in real time. In these contexts, one can imagine that our proposed models can hold natural relevance, as we have motivated their construction at the level of the patient's health progression. Drawing inspiration from problems such as in \citet{shi2021timing}, let us now numerically demonstrate how this modeling framework may be used in real-time healthcare decision making settings.}

\edit{One such critical issue may arise when a remotely hospitalized patient deteriorates and reaches their call-in threshold, but due to stochasticity, there are no available resources on-site for admission. 
A reasonable policy in this case would be to send an on-site patient home to complete their treatment remotely. In other words, this patient will be swapped with the remotely hospitalized patient who was called in. Notice that this  ``swapping problem'' effectively becomes a real-time addendum to the design decisions in this paper.}

\edit{There could be different ways to select the on-site patient who will be swapped. A straightforward approach would be to swap the patient with the best current health score. However, our model allows us to propose a more effective dynamic policy. This policy selects the patient who, based on their current state, incurs the minimum expected cost to complete treatment remotely. Additionally, we factor in the associated cost of potential deterioration and the likelihood of being called back into the hospital.}

\edit{Let $\mathcal{J}$ denote the set of on-site patients eligible to be swapped and complete their hospitalization remotely (perhaps by being close enough to discharge). For each $j\in\mathcal{J}$, let $k_j$ represent the patient type and $\tilde{s}_j$ denote the current health score. 
With a slight abuse of notation, we denote the parameters of each patient $j$ by a superscript $j$; for example, $T^j$ (instead of $T^{k_j}$) represents the distance of patient $j$ from the hospital. Lastly, we define $s_j := \tilde{s}_j + \theta_T^j T^j$ as the patient's expected health score after traveling home.}

\edit{In a similar way as we constructed the objective function in (\ref{eq:obj_func_linear}), we now calculate the following average cost for each on-site patient $j\in\mathcal{J}$ whenever a swap is required:
\begin{align*}
\mathcal{C}_j\left(s_j\right) & = 
h_R \Ex \left[\tau_{R}(s_j,a_C^j)   \right] 
+ h_T^j T^j + 
\big(h_T^j T^j +  \Ex \left[\tau_{H}(s_j,a_C^j,Z^j)   \right] \big) p_{s_j}\left(a_C^j\right)
\\
& = \frac{h_R^j}{\theta_R^j}\left(\left(1-p_{s_j}\left(a_C^j\right)\right)s_j - p_{s_j}
\left(a_C^j\right)a_C^j\right) + h_T^j T^j+ \left(h_T^j T^j + \frac{h_H^j}{\theta_H^j}\left(a_C^j+s_j+\theta_T^j T^j\right) \right)p_{s_j}\left(a_C^j\right). 
\end{align*}
Here, the first term represents the average remaining LOS remotely (the average time to hit either zero or the call in threshold when starting at $s_j$), the second term is the traveling cost from the hospital to the patient's home, and the third term represents the hospitalization cost on-site in case of deterioration.}

\edit{The dynamic policy we suggest is an index-based policy. Under
such a policy, the patient selected for a swap is simply chosen from those eligible according to the minimal cost incurred to complete their hospitalization at home. In particular, such a policy states that the patient $j_s$ chosen for discharge satisfies $
j_s = \argmin_{j\in\mathcal{J}} \mathcal{C}_j\left(s_j\right)$. 
Implementing such a policy requires data about each individual on-site patient, which hospitals typically collect consistently. Ensuring that a patient is not swapped more than once can be easily managed by constructing $\mathcal{J}$ so that it includes only patients who have not been previously swapped.}

\edit{To demonstrate the benefit of this modeling framework and the individual-level approach of this model, we will now use a stochastic simulation model to evaluate its performance and compare it with a policy that swaps the patient with the best (lowest) current health score. The model includes two types of patients arriving according to a Poisson process. The call-in threshold for each patient type in each scenario were set based on the optimal system design derived in Section \ref{sec: Multiple Item Types} for a given baseline parameter set.}



\edit{Recall that our dynamic policy is activated when a remote patient requires hospital admission, but no resources are available due to stochasticity. In this scenario, we initiate a swap between one of the on-site patients and the remotely deteriorating patient who needs on-site care. As a benchmark policy, we use an intuitive practice: swapping the on-site patient with the best health score. We refer to this policy as Policy 1. Our suggested policy, denoted Policy 2, is instead based on the cost-index $\mathcal{C}_j\left(s_j\right)$.} 

\begin{table}[h]
\caption{\edit{Comparing the long-run average cost for two patient types under Policy 1 and Policy 2. {$\bar{S}^1=\bar{S}^2=15$, $\sigma_R^1 = \sigma_R^2=1$, $\theta_T^1=\theta_T^2=0.1$. In Scenarios 1a--1c, $\lambda^1=\lambda^2=2$,  $(\theta_R^1,\theta_R^2) = (0.2,0.5)$, $h_R^1=h_R^2 = 450$; in Scenarios 2a--2c, $\lambda^1=\lambda^2=4$,  $(\theta_R^1,\theta_R^2) = (0.2,0.6)$, $(h_R^1,h_R^2) = (420,1530)$; in Scenarios 3a--3c, $\lambda^1=\lambda^2=3$, $(\theta_R^1,\theta_R^2) = (0.2,0.6)$, $h_R^1=h_R^2 = 450$; in Scenarios 4a--4c, $\lambda^1=\lambda^2=3.5$, $(\theta_R^1,\theta_R^2) = (0.2,0.5)$, $h_R^1=h_R^2 = 450$. The long-run average is estimated using a time horizon of $10^4$.}}\\}
\begin{center}
\label{table: dynamic policy comparison}
\begingroup
\edit{\small
\begin{tabular}{  c | c | c | c |  c  c  c  c  c  } 
Scenario & ($\gamma_1,\gamma_2$) & $C$ & ($h_T^1 T^1/h_T^2T^2$) & ($\theta_H^1,\theta_H^2$) & ($\sigma_R^1,\sigma_R^2$) & Policy 1   & Policy 2  & Improvement\\ \hline\hline
1a   & \multirow{3}{*}{$\gamma_1>0,~ \gamma_2>0$}  & \multirow{3}{*}{15}  & \multirow{3}{*}{1:2}  & (0.143, 0.25) & (1, 1) & 6,995	& 6,756	 & 3.42\%  \\ \cline{1-1}\cline{5-9}
1b & & & & (0.0715, 0.125)  &  (1, 1) & 9,252 & 7,226 &21.9\% \\\cline{1-1}\cline{5-9}
1c & & & & (0.143, 0.25)  &  (1, 1.2) & 9,828 & 8,232 &16.24\% \\\cline{1-1}\cline{5-9}\hline\hline
2a   & \multirow{3}{*}{$\gamma_1<0,~ \gamma_2<0$}  & \multirow{3}{*}{25}  & \multirow{3}{*}{4:1}  & (0.5, 0.5)  & (1, 1) & 19,761	& 19,044	 & 3.63\%  \\ \cline{1-1}\cline{5-9}
2b & & & & (0.25, 0.25)  &  (1, 1) & 23,364 & 9,767 & 15.4\% \\\cline{1-1}\cline{5-9}
2c & & & & (0.5, 0.5)  &  (1, 1.2) & 23,784 & 20,299 & 14.65\% \\\cline{1-1}\cline{5-9}\hline\hline
3a   & \multirow{3}{*}{$\gamma_1>0,~ \gamma_2<0$}  & \multirow{3}{*}{20}  & \multirow{3}{*}{3:1}  & (0.143, 0.5)  & (1, 1) & 16,182	& 15,438	 & 4.6\%  \\ \cline{1-1}\cline{5-9}
3b & & & & (0.0715, 0.25)  &  (1, 1.2) & 20,654 & 17,039 & 17.5\% \\\cline{1-1}\cline{5-9}
3c & & & & (0.5, 0.5)  &  (1, 1.2) & 21,408 & 18,111 & 15.4\% \\\cline{1-1}\cline{5-9}\hline\hline
4a   & \multirow{3}{*}{$\gamma_1<0,~ \gamma_2>0$}  & \multirow{3}{*}{20}  & \multirow{3}{*}{1:4} & (0.5, 0.25) & (1, 1) & 16,751	& 16,235 & 3.08\%  \\\cline{1-1}\cline{5-9}
4b & & & & (0.25, 0.125)  &  (1, 1) & 30,875 & 26,347 & 14.67\% \\\cline{1-1}\cline{5-9}
4c & & & & (0.5, 0.25)  &  (1, 1.2) & 19,882 & 17,089 & 14.05\% \\\cline{1-1}\cline{5-9}\hline\hline
\end{tabular}}
\endgroup
\end{center}
\end{table}

\edit{Our experiments, summarized in Table \ref{table: dynamic policy comparison}, include twelve scenarios grouped into three categories. The hospitalization costs selected are  consistent with the hybrid hospital model examined in \cite{zychlinski2024tele}. Scenario ``a” in each group simulates the system using parameters for which the optimal thresholds were initially derived. Because the system operates according to its optimal design, the number of swaps is small, resulting in a relatively small cost difference between the policies, with our proposed Policy 2 outperforming the standard practice.}
\edit{Scenario ``b'' in each group simulates the system with reduced hospital recovery rates, $\theta_H$. This reduction effectively describes a situation where patients have longer hospital stays. Consequently, there are more swaps, which amplify the cost difference between the policies.}
\edit{Finally, Scenario ``c'' in each group simulates the system with increased variability in remote recovery, leading to a higher call-in probability and more swaps. In this scenario, too, our proposed policy outperforms the standard practice. Note that the cost differences between the policies could be significantly greater if the disparity between the two classes, in terms of distance from the hospital and travel costs, were to increase.} 

\edit{As is suggested by the dominance of our proposed policy in these simulation experiments, our severity-modeling framework may hold significant potential practically for these second-order real-time control problems that complement the first-order design and static planning problems we have analyzed in this paper. In the interest of brevity, we reserve formal analysis of these online decisions for future research.}


\section{Discussion and Conclusion}
\label{sec:Conclusions and Direction for Future Research}
The hybrid hospital model constitutes a service network design problem. The decision of whether to admit a patient remotely or on-site entails the efficient allocation of resources across the two hospitalization modes. To address this, we adopt a modeling approach that captures the dynamic progression of individual health conditions within the network and during travel. System design optimization, in this context, revolves around establishing the call-in threshold that minimizes the total operational costs, consequently influencing the optimal resource allocation \edit{between on-site and at-home hospitalization and among different patient types}. 

Managerially, our results both offer guidance on how hospitals should allocate resources between on-site and at-home care, and identify a potential cautionary tale, in that distant patients may not actually be best served by remote hospitalization. \edit{The main results of this paper, Theorems~\ref{theorem: T properties},~\ref{theorem:opt_a_cap}, and~\ref{theorem: multi type sol char}, share an analytical through-line that offers an immediate managerial ``spot check'' which depends on just three values and may actually be best summarized by \eqref{gammaGammaEq}. Given abundant resources, the difference of marginal costs, $\gamma$, portends the range and shape of the optimal call-in threshold for home hospitalization. Then, under limited resources, the constrained optimal call-in threshold becomes equivalent to an unconstrained optimal level in which $h_H$ and $h_R$ are both shifted by an amount $\Gamma \geq 0$. By~\eqref{gammaGammaEq}, the way that $\Gamma$ effects $\gamma$ will depend solely on the comparison of the corresponding recovery rates, $\theta_R$ and $\theta_H$. In the most realistic setting where recovery is faster on-site rather than at home ($\theta_R < \theta_H$), \eqref{gammaGammaEq} shows that the range of patient health conditions and distances viably served by home hospitalization shrinks even further when resources are scarce. Finally, by Theorem~\ref{theorem: multi type sol char}, this same underlying phenomenon is immediately carried through to the model's most faithful representation of reality, where the same general shapes and shrinking occur when multiple types of patients must vie for limited medical resources in the hybrid health network.}

Qualitatively, these \edit{structural} insights may also be relevant in other parts of public life, e.g.,~if online education is considered as an alternative for a rural school with declining enrollment. As in the case of remote hospitalization, these challenges may be further heightened  by inequities of internet access for those who live in rural areas \citep{lai2021revisiting}. Much like we have discussed for the pitfalls of relying on remote hospitalization to serve rural communities, our results \edit{may caution} towards potential over-reliance on online education, in which recent data from the pandemic has revealed stark and concerning disparities in rates of learning relative to just before COVID-19 began \citep{halloran2021pandemic,goldhaber2022consequences}.

Whether in the focal hospitalization application or in other relevant areas, one possible limitation of our model is the underlying assumption that all patients eventually recover. Indeed, the negative drifts of the Brownian motions naturally imply that every patient's severity will hit zero in finite time almost surely. This presents a natural opportunity to generalize and model mortality or another form of negative outcome. While the assumption of guaranteed recovery may be conservative, we believe that this actually emphasizes both the importance of careful hybrid hospital design and  the fragility of the relationship between the remote format and patience distance. That is, viewing our results with  the eventual recovery assumption in mind, we see that even when the worst that can happen is \emph{added cost}, remote hospitalization is still only viable for a limited range of patient distances (which may be further limited by the initial health severities), even when the operations are designed optimally as we describe.

There are several additional future research directions, \edit{as exemplified by Section~\ref{subsec: A Dynamic Policy}}. 
\edit{In particular, our simulation experiments show that this model offers} a foundational framework for the development of dynamic control policies. The call-in threshold establishes a reference point that can be further refined through real-time performance enhancements. Ultimately, the optimal control strategy would introduce state-dependent call-in decisions, which may be adapted based on perturbations from the initial optimal baseline decisions.

\bibliographystyle{informs2014} 
\bibliography{Bibliography.bib} 



\newpage

\begin{APPENDICES}
\renewcommand{\theHchapter}{A\arabic{chapter}}


\section{Proofs}
\label{appendix: Proofs}

\paragraph{\bf Proof of Lemma \ref{lemma:W_on_a}.}

Recall that $p_x(a)$ is the hitting probability given by
\[
p_x(a) :=\mathbb{P}\left(\calB^{R}(\tau_{R}(x,a))=a+x\right)= \frac{1-e^{-\rho x}}{e^{\rho a} - e^{-\rho x}},
\]
and therefore, its first derivative with respect to $a$ is
\[
p_x'(a)=-\frac{\rho e^{\rho a}}{e^{\rho a} - e^{-\rho x}}p_x(a)<0.
\]

Our goal is to prove that $W'_H(a)<0$ and $W'_R(a)>0$. We begin with $W_H$. Recall that:
\[
W_H(a) = \frac{\lambda p_x(a)}{\theta_H} \left(a+x+T\theta_T\right).
\]
Therefore and since $p_x'(a)\neq0$, 
\begin{align}\label{eq:derivative_W_H}
W_H'(a) = \frac{\lambda}{\theta_H}\left(p_x'(a)(a+x+\theta_T T)+p_x(a)\right)=\frac{\lambda}{\theta_H}p_x'(a)\left(a+x+\theta_T T+\frac{p_x(a)}{p_x'(a)}\right).
\end{align}
Now,
\begin{align}\label{eq:p_p'}
\frac{p_x(a)}{p_x'(a)} = -\frac{e^{\rho a}-e^{-\rho x}}{\rho e^{\rho a}} = -\frac{1}{\rho}(1-e^{-\rho(a+x)}) > -(a+x),
\end{align}
where the inequality is because $1-e^{-x}< x$  for $x>0$. Therefore, 
\begin{align}\label{a_lem_1_1}
    a+x+\theta_T T+\frac{p_x(a)}{p_x'(a)}> \theta_T T ~~ \Rightarrow~~ a+x+\frac{p_x(a)}{p_x'(a)} >0.
\end{align}
Multiplying both sides by $\frac{\lambda}{\theta_H}p_x'(a)$, the result follows since $p_x'(a)<0$. We turn to $W_R(a)$. We have:
\[
W_R(a) = \frac{\lambda}{\theta_R}\left(\left(1-p_x(a)\right)x - p_x(a)a\right)= \frac{\lambda}{\theta_R}x-\frac{\lambda}{\theta_R}p_x(a)(a+x),
\]
and therefore (and again, because $p_x'(a)\neq0$),
\begin{align}\label{eq:derivative_W_R}
    W_R'(a)=-\frac{\lambda}{\theta_R}(p_x'(a)(a+x)+p_x(a)) = -\frac{\lambda}{\theta_R}p_x'(a)\left(a+x+\frac{p_x(a)}{p_x'(a)}\right)>0,
\end{align}
where the inequality is from \eqref{a_lem_1_1} and since $p_x'(a)<0$. \eProof

\paragraph{\bf Proof of Proposition \ref{prop:W_total_on_a}.}
Recall that $W_T(a)=W_H(a)+W_R(a)$ and that we wish to characterize the dependence of $W_T$ on $a$. For ease of notation, denote $r=\theta_H/\theta_R$. We have:
\begin{align*}
W_T'(a) = W_H'(a) + W_R'(a)&\overset{\eqref{eq:derivative_W_H},\eqref{eq:derivative_W_R}}{=}\frac{\lambda}{\theta_H}p_x'(a)\left(a+x+\theta_T T+\frac{p_x(a)}{p'_x(a)}\right)-\frac{\lambda}{\theta_R}p'_x(a)\left(a+x+\frac{p_x(a)}{p'_x(a)}\right)\cr
   &\overset{\eqref{eq:def_of_Delta}}{=} \frac{\lambda p'_x(a)}{\theta_H} \left((1-r)\left(a+x+\frac{p_x(a)}{p'_x(a)}\right)+\theta_T T\right)\cr
   &\overset{\eqref{eq:p_p'}}{=}\frac{\lambda |p'_x(a)|}{\rho\theta_H}\left((1-r)(1-\rho(a+x)-e^{-\rho(a+x)})-\theta_T T \rho\right), 
\end{align*}
where the \edit{absolute value is used} because $p'_x(a)<0$. 

We start with Case 1, where $\theta_H/\theta_R\leq 1 \iff 1-r \geq 0$. 
Using the inequality $1-e^{-x}\leq x$, we get that $1-\rho(a+x)-e^{-\rho(a+x)}\leq 0$. Thus, if $1-r \geq 0$, then $W'_T(a)<0$.

We turn to Case 3. Note that when $r>1$, $1-r=-|1-r|$. In this case,
\begin{align*}
W_T'(a)\overset{\text{if }r>1}{=}\frac{\lambda |p'_x(a)||1-r|}{\rho}\left(-1+\rho(a+x)+e^{-\rho(a+x)}-\frac{\theta_T T \rho}{|1-r|}\right).
\end{align*}

Denoting $h(a):= -1+\rho(a+x)+e^{-\rho(a+x)}$, we have:
\begin{align}\label{eq:properties_of_h}
    &h(0)= \rho x-1+e^{-\rho x}{>}0~~~~~\text{because } e^{-x}>1- x  \text{ for } x>0,\cr
    &h'(a)=\rho-\rho e^{-\rho(a+x)}=\rho(1- e^{-\rho(a+x)})>0,\cr
    &h(a)\xrightarrow{a\rightarrow \infty}\infty.
\end{align}

Thus, $W_T'(a)$ can be negative, if and only if its value at $a=0$ is negative. Clearly, this \textit{does not} happen if $h(0)\geq \frac{\theta_T T \rho}{ |1-r|}$, or, written differently, if 
$$|1-r|\geq \theta_T T \rho\left(  \rho x-1+e^{-\rho x}\right)^{-1}\overset{\text{if }r>1}{\iff}r\geq 1+\Delta,$$
where $\Delta$ is defined in \eqref{eq:def_of_Delta}.

Lastly, we turn to Case 2 where $1<r<1+\Delta$. This implies that $h(0)<\frac{\theta_T T \rho}{ |1-r|}$. In this case, by \eqref{eq:properties_of_h}, it is clear that $W_T'(0)<0$, and that there exists $a_0>0$ such that $W_T'(a)<0$ for $a \in [0,a_0)$, $W_T'(a_0)=0$, and  $W_T'(a)>0$ for $a \in (a_0,\infty)$. This concludes the proof.
\eProof

\paragraph{\bf Proof of Proposition \ref{prop:feasibility_shared}.}

Per section \ref{subsec:Feasibility_Region}, the feasibility region of  optimization problem \eqref{eq:opt_RSharing} is
\[
{\calC_{FR}} = \{(\lambda,C) \in \mathbb{R}_+^2:  W_T(a_{\min})\leq C\}.
\]
All that is left is to characterize $a_{\min}$. By Proposition \ref{prop:W_total_on_a}, if $\theta_H/\theta_R\leq 1$, then $W_T(a)$ is strictly \tb{decreasing} in $a$. Thus, in this case, $a_{\min}=0$ which proves the first item.
Again by Proposition \ref{prop:W_total_on_a}, if $\theta_H/\theta_R \geq 1+\Delta$, then $W_T(a)$ is strictly \tb{increasing} in $a$. Thus, in this case, $a_{\min}=\bar{S}-x-T\theta_T$, which proves the second item.

Finally, in the case where $1<\theta_H/\theta_R<1+\Delta$, by Proposition \ref{prop:W_total_on_a}, there exists $a_0>0$ such that $W_T'(a)<0$ for $a \in [0,a_0)$, $W_T'(a_0)=0$, and  $W_T'(a)>0$ for $a \in (a_0,\infty)$. If $a_0<\bar{S}-x-T\theta_T$, then $a_{\min}=a_0$. Otherwise, $\bar{S}-x-T\theta_T \in (0,a_0]$, and since $W_T(a)$ is strictly decreasing in this interval, we have $a_{\min}=\bar{S}-x-T\theta_T $ which completes the proof of the third item and the proposition. 
\eProof

\paragraph{\bf Proof of Proposition \ref{prop:opt_a_New2}.}
We first provide technical characterizations of the objective function $V(a)$. 
\begin{lemma}\label{lemma: prop V}
The value function $V(a)$ satisfies the following:

If $\gamma\geq0$, then $V(a)$ is strictly decreasing in $a \in\mathbb{R}_+$; else ($\gamma<0$),  
    \begin{itemize}
        \item If $\beta\geq0$, or $\gamma(1 - e^{-\rho x})/\rho < \beta < 0$, then $V(a)$ is unimodal with a unique minimum over $a \in\mathbb{R}_+$.
        \item If $\beta \leq \gamma(1 - e^{-\rho x})/\rho$, then $V(a)$ is strictly increasing in $a \in\mathbb{R}_+$.
    \end{itemize}
\end{lemma}

From Lemma~\ref{lemma: prop V}, $\gamma \geq 0$ yields that $V(a)$ is strictly decreasing, and thus the largest allowable threshold is optimal: $a^* = \bar{A}$.
Then, if $\gamma < 0$ and $\beta \geq \gamma(1-e^{-\rho x})/\rho$, then Lemma~\ref{lemma: prop V} provides that there is a unique optimal solution. Moreover, because the definition of $\tilde a$ in Equation~\eqref{tildeAEq} is precisely the first order condition in Equation~\eqref{aFOC} within the proof of Lemma~\ref{lemma: prop V}, we can see that $\tilde a$ is the unique maximizer of $V(a)$. If $\tilde a$ is within the maximum allowable threshold size, then it is optimal for the unlimited capacity problem, but if $\tilde a > \bar{A}$, then we can see that $V'(a) > 0$ for all $a \in [0, \bar{A}]$, meaning $\bar{A}$ is optimal. 
 Hence,  $a^* = (\tilde a \wedge \bar{A})$ is the optimal threshold.
Finally, for the remaining case and again by Lemma~\ref{lemma: prop V}, if $\beta \leq \gamma (1 - e^{-\rho x})/\rho$, then $V(a)$ is always increasing, and thus the optimal threshold is as low as possible, directing all patients immediately to on-site hospitalization: $a^* = 0$.

To complete the proof, let us verify that $\tilde a$ given by the solution in Equation~\eqref{solW} is indeed positive and obtained by the principal branch of the Lambert-W function. Note that the existence of a unique, positive  solution to first order condition in Equation~\eqref{aFOC} is already guaranteed for $\gamma < 0$ and $\beta > \gamma (1 - e^{-\rho x}) / \rho$ through the preceding linear-and-exponential-function arguments; the focus now is simply on proving the correctness of Equation~\eqref{solW}. Rearranging~\eqref{aFOC} and multiplying both sides by $e^{\beta \rho / \gamma - 1}$, we have that $\tilde a$ will be the $a$ that solves
$$
-
e^{-\rho x + \beta \rho / \gamma - 1}
=
\left(
\rho a + \beta \rho / \gamma - 1
\right)
e^{\rho a + \beta \rho / \gamma - 1}
.
$$
Before further manipulating this equation, let us inspect the terms in the exponent on the left-hand side. If $\beta \geq 0$, it is clear that $-\rho x + \beta \rho / \gamma - 1 < 0$, so let us focus on $\gamma (1 - e^{-\rho x}) / \rho < \beta < 0$. Dividing by $\gamma/\rho < 0$, we have that $0 < \beta \rho / \gamma < 1 - e^{-\rho x}$, and, furthermore, by adding $-1 -\rho x$ to each side, we have that $-\rho x + \beta \rho / \gamma - 1 < - \rho x - e^{-\rho x} < 0$. Hence, for all $\beta > \gamma ( 1 - e^{-\rho x} ) / \rho$,  $-
e^{-\rho x + \beta \rho / \gamma - 1} \in (-1/e, 0)$.

For the identity $\mathsf{W}(z e^{z}) = z$ to hold on the principal branch of the Lambert-W, we must have $z \geq - 1$. Hence, as a final step, let us show that $\rho \tilde a + \beta \rho / \gamma - 1 \geq -1$. If $\beta \leq 0$, this is immediately true by the fact that $\rho > 0$, $\tilde a > 0$, and $\gamma < 0$, so let us focus on the $\beta > 0$ case. If $\tilde a > -\beta / \gamma$, then we can apply the Lambert-W principal branch identity, and Equation~\eqref{solW} will follow immediately. To see that this is indeed true, we return to the linear-and-exponential-function arguments. Notice that, at $a = -\beta/\gamma > 0$, the left-hand side of Equation~\eqref{aFOC} is $e^{\rho \beta / \gamma} < 1$, whereas the right-hand side simplifies to $e^{\rho x} > 1$. Therefore, the linear function has not yet crossed the exponential function, implying $\tilde a > -\beta / \gamma$.\eProof

\paragraph{\bf Proof of Lemma \ref{lemma: prop V}.} 
To begin, let us obtain a first order condition for $V(a)$. The derivative of the cost function with respect to $a$ is
$$
V'(a)
=
\beta 
p'_x(a)
+
\gamma a 
p'_x(a)
+ 
\gamma p_x(a)
.
$$
Since $p'_x(a)< 0$, the cost derivative simplifies to
\begin{align}
V'(a)&
= p'_x(a)\left(\beta+\gamma a+\gamma \frac{p_x(a)}{p'_x(a)}\right)\overset{\eqref{eq:p_p'}}{=}p'_x(a)(\beta+\gamma a-\gamma\frac{1}{\rho}(1-e^{-\rho(a+x)}))\cr
& = \frac{|p'_x(a)|}{\rho}\left(\gamma(1 - e^{-\rho (x + a)})-\beta \rho-\gamma \rho a\right).
\label{dVdef}
\end{align}
Since $\rho>0$, and given that $x > 0$, $|p'_x(a)|$ is strictly positive for all $a \geq 0$, the sign of $d V / d a = 0$ matches the sign of $\gamma(1 - e^{-\rho (x + a)})-\beta \rho-\gamma \rho a$. We can see that the $a$-derivative of this expression is
\begin{align}
\left(
\gamma(1 - e^{-\rho (x + a)})
-
\rho \beta
-
\gamma \rho a
\right)'
=
-
\gamma \rho \left(1 - e^{-\rho(x+a)}\right)
,
\label{dVsign}
\end{align}
and thus we can recognize that whether or not $V'(a)$ will be 0 for some $a \in \mathbb{R}_+$ purely depends on the sign of $\gamma$ and the initial sign of $V'(a)$ at $a = 0$. Note that this does \emph{not} necessarily imply convexity or concavity: $V''(a)$ need not match $\gamma$ in sign. Hence, $V'(a)$ may fluctuate between increases and decreases across values of $a \in \mathbb{R}_+$, but it will cross 0 at most once on this range.

This leads us to consider when $V'(a) = 0$. Rearranging $\gamma(1 - e^{-\rho (x + a)})
-
\rho \beta
-
\gamma \rho a$, we find the following first order condition: $a$ is a candidate optimal threshold solution, if and only if
\begin{align}
e^{-\rho a}
=
\left(
1 - \rho \beta  / \gamma - \rho a
\right)
e^{\rho x}
\label{aFOC}
.
\end{align}
Now, let us notice that, as functions of $a$, the left-hand side of Equation~\eqref{aFOC} is a decaying exponential (exponential with negative rate $- \rho < 0$) and right-hand side is simply a linear function with  slope $-\rho e^{\rho x} < - \rho$. Hence, the right-hand side function will intersect the left-hand side function at most once on $a \in \mathbb{R}_+$. To evaluate where this occurs, let us proceed case wise. 

Beginning with $\gamma > 0$, we can see that, by definition, this implies that $\beta > 0$ also. Furthermore, the definitions of $\beta$ and $\gamma$ also reveal that
$$
\frac{\beta}{\gamma}
=
x 
+ 
\frac{1}{\gamma}
\left(
h_T T
+
\frac{h_H \theta_T T}{\theta_H}
\right)
>
x
,
$$
and thus we have that
$$
\left(1 - \frac{\rho \beta}{\gamma}\right) e^{\rho x}
<
(1 - \rho x) e^{\rho x}
\leq 
1
.
$$
Therefore, the left-hand side of Equation~\eqref{aFOC} at $a= 0$ is strictly greater than the right-hand side of~\eqref{aFOC} at $a = 0$, implying that, respectively, this exponential function is always above the negative slope line, and thus there is no solution to the first order condition in this setting. By applying these arguments to Equation~\eqref{dVdef} and recalling that $\gamma > 0$ in this case, Equation~\eqref{dVsign} then shows that $V'(a)<0$ for all $a \in \mathbb{R}_+$. For $\gamma = 0$, we can quickly recognize from Equation~\eqref{dVdef} that, again, $V'(a)<0$ for all $a$.

Let us now suppose that $\gamma < 0$. Through Equation~\eqref{dVsign}, we have that, once $V'(a)>0$, it will remain positive for all increasing values of $a$. So, we now partition the $\gamma < 0$ case into sub-cases evaluating the initial sign of $V'(a)$ at $a = 0$. Here, we see that, at $a = 0$, $\gamma(1 - e^{-\rho (x + a)})
-
\rho \beta
-
\gamma \rho a
=
\gamma(1 - e^{-\rho x})
-
\rho \beta
.
$
In sub-case that $\beta \geq 0$, or, equivalently, $-\left( h_T T + h_H \theta_T T / \theta_H \right) / x \leq \gamma < 0$, we find that 
$$
\gamma(1 - e^{-\rho x})
-
\rho \beta
<
0
,
$$
and thus $V(a)$ is decreasing at $a = 0$.
This also implies that the right-hand side of Equation~\eqref{aFOC} starts above the exponential in the left-hand side of~\eqref{aFOC}, ensuring that there will be a unique solution to the first order condition on $\mathbb{R}_+$. Similarly, if $\beta < 0$ but $\gamma(1 - e^{-\rho x}) < \rho \beta$ still holds, then the same arguments apply.

Finally, if $\beta \leq \gamma(1 - e^{-\rho x})/\rho$ with $\gamma < 0$, then $V'(a) \geq 0$ at $a = 0$, and, by Equation~\eqref{dVsign}, it will remain so for all $a \in \mathbb{R}_+$. \eProof

\paragraph{\bf Proof of Theorem~\ref{theorem: T properties}.}
We begin by proving the first statement: $a_\infty^* > 0$ if and only if $T_{LB} < T < T_{UB}$ under the case that $\gamma  \geq 0$. If $\gamma  \geq 0$, then by Proposition~\ref{prop:opt_a_New2}, $a^*_\infty = \bar A = \bar S - x - T \theta_T$. Hence, it is immediately true that $a_\infty^* > 0$ if and only if $T < T_{UB}$. Since $T_{LB} \leq 0$ by consequence of $\gamma \geq 0$, we complete the proof in this setting. 

If $\gamma < 0$, Proposition~\ref{prop:opt_a_New2} provides that $a_\infty^* = (\tilde a \wedge \bar A)$ if $\beta > \gamma(1- e^{-\rho x})/\rho$, where $\tilde a > 0$ is given by Equation~\eqref{solW}. By the preceding arguments, notice that if and only if $T \geq T_{UB}$, then $\bar A = 0$. Now, we can further observe that among the streamlined model coefficients, $\alpha$, $\beta$, $\gamma$, only $\beta$ depends on $T$. Specifically, with the additionally defined $\eta$, we have that $\beta = \gamma x + \eta T$. Hence, the condition for $\tilde a > 0$ can be re-expressed to
$$
\gamma x + \eta T > \gamma(1- e^{-\rho x})/\rho
,
$$
and this immediately simplifies to $T > T_{LB}$. Hence, we have that $\tilde a > 0$ if and only if $T > T_{LB}$ and that $\bar A > 0$ if and only if $T < T_{UB}$, which proves that $a_\infty^* > 0$ if and only if $T \in (T_{LB}, T_{UB})$. In particular, $a_\infty^* = (\tilde a \wedge \bar A) > 0$ on this interval. Moreover, let us observe that the argument of the Lambert-W function in the expression for $\tilde a$ in~\eqref{solW} simplifies to
$$
-e^{-\rho x + \frac{\beta \rho }{ \gamma } - 1}
=
-e^{-\rho x + \frac{\rho}{\gamma}\left( \eta T + \gamma x \right) - 1}
=
-e^{\frac{\eta \rho}{\gamma} T - 1}
.
$$
Likewise, Equation~\eqref{solW} itself simplifies to
\begin{align}
x + \tilde a
=
\frac{1}{\rho}
\left(
1
+
\mathsf{W}\left(
-e^{\frac{\eta \rho}{\gamma} T - 1}
\right)
\right)
-
\frac{\eta}{\gamma} T
.
\label{solWbyT}
\end{align}

Considering each of the two components of $(\tilde a \wedge \bar A)$ individually, let us observe how they each depend on $T$. Starting with $\tilde a$, by Equation~\eqref{solWbyT}, we can see that
\begin{align*}
\frac{\partial \tilde a}{\partial T}
&=
\frac{1}{\rho}
\frac{\partial}{\partial T}
W\left(
-
e^{\rho \eta T / \gamma - 1}
\right)
-
\frac{\eta}{\gamma}
.
\end{align*}
Using the fact that $\mathrm{d}W(z)/\mathrm{d}z = W(z)/(z(1+W(z)))$ for $z \in (-1/e, 0)$, this simplifies to
\begin{align*}
\frac{\partial \tilde a}{\partial T}
&=
-
\frac{\eta}
{\gamma}
\frac{
1
-
W\left(
-
e^{\rho \eta T / \gamma - 1}
\right)
}
{
1
+
W\left(
-
e^{\rho \eta T / \gamma - 1}
\right)
}
.
\end{align*}
Because $\gamma < 0$ and because the principal branch Lambert-W function is greater than $-1$ for all arguments greater than $-1/e$, we have that $\partial \tilde a / \partial T > 0$ for all values of $T$. Turning to the second component within the minimum, we can quickly observe from the definition of $\bar A$ that
$$
\frac{\partial \bar A}{\partial T}
=
- \theta_T
.
$$

Thus, the dependence of $a_\infty^*$ on $T$ is clear: starting from $T_{LB}$, $a_\infty^*$ increases according to $\tilde a$ until $\tilde a$ intersects $\bar A$, and then decreases from this point until reaching $T_{UB}$. Hence, we can find that this change point is given by the unique $T$ at which $\tilde a = \bar A$. Setting the two quantities equal to one another, we have
\begin{align*}
\frac{1}{\rho}
\left(
1
+
W\left(
-e^{\frac{\eta \rho}{\gamma} T - 1}
\right)
\right)
-
\frac{\eta}{\gamma} T
-
x
&=
\bar S - x - T \theta_T
,
\end{align*}
 and this simplifies to the definition of $\hat T$ in Equation~\eqref{ThatDef}.
\eProof

\paragraph{\bf Proof of Theorem \ref{theorem:opt_a_cap}.} 
Recall the following notation and previously proven results
\begin{enumerate}
    \item $a_{\infty}^*:=\argmin_{a \in \mathcal{A} } V(a)$
    \item $a_{\min}:=\argmin_{a \in \mathcal{A} } W_T(a)$
    \item Both $a_{\infty}^*$ and $a_{\min}$ are unique.
    \item Based on the analysis in the proof of Proposition \ref{prop:W_total_on_a}, depending on the problem parameters, there are 3 possible ways $W_T(a)$ behaves as a function of $a$. 
    \begin{enumerate}
        \item $W_T(a)$ is strictly increasing, then $a_{\min}=0$. Importantly and in particular, $W_T(a)$ is strictly increasing to the right of $a_{\min}$.
        \item $W_T(a)$ is strictly decreasing, then $a_{\min}=\bar{A}$. Importantly and in particular, $W_T(a)$ is strictly decreasing to the left of $a_{\min}$. Meaning, as we decrease $a$, starting from $a_{\min}$, the value of $W_T(a)$ increases.
        \item $W_T(a)$ has a unique minimum in $(0,\bar{A})$, it strictly decreases before it and strictly increases after.
    \end{enumerate}
    \item The conclusion from the item above is that if we pick any $\hat{a} \in \mathcal{A}$ which satisfies $\hat{a}\neq a_{\min}$ (but both $\hat{a}> a_{\min}$ and $\hat{a}< a_{\min}$ are possible), then if we move from $\hat{a}$ to $a_{\min}$, the value of $W_T(a)$ \textbf{strictly decreases}.
    \item Based on the analysis in the proof of Proposition \ref{prop:opt_a_New2}, depending on the problem parameters, there are 3 possible ways $V(a)$ behaves as a function of $a$. 
    \begin{enumerate}
        \item $V(a)$ is strictly increasing.
        \item $V(a)$ is strictly decreasing
        \item $V(a)$ decreases, then has a unique minimum in $\mathbb{R}_+$, then strictly increases.
    \end{enumerate}
    \item From the last item, we can conclude that if we move from $a_{\edit{\infty}}^*$ to any other $\hat{a} \in \mathcal{A}$, $V(a)$ \textbf{strictly increases}. We can also deduce that $V'(a)$ can be zero at most once, and that if it does, then this point is a minimum. 
\end{enumerate}

First, if $W_T(a_{\min})=C$, since $a_{\min}$ is unique, $a_{\min}$ is the only feasible value for $a$ in $\mathcal{A}$, and therefore it is the unique solution, i.e., $a_{C}^*=a_{\min}$.
Next, assume that $W_T(a_{\min})<C$. If $W_T(a_{\infty}^*)\leq C$, then $a_{\infty}^*$ is feasible and uniquely minimizes $V(a)$ in $\mathcal{A}$. Thus it is the unique solution, i.e., $a_{C}^*=a_{\infty}^*$.

We are left with the case where $W_T(a_{\min})<C$ and $W_T(a_{\infty}^*)> C$. In particular, we must have $a_{\min} \neq a_{\infty}^*$. By the properties listed above, when we start at $a_{\infty}^*$ and go towards $a_{\min}$, $W_T(a)$ must strictly decrease and $V(a)$ must strictly increase. Since $W_T(a)$ is continuous, there must be a value for $a$, call it $\hat{a}$, strictly between $a_{\infty}^*$ and $a_{\min}$ for which $W_T(\hat{a})=C$, which also means $\hat{a}$ is feasible. Additionally, any other value for $a$ before we reach $\hat{a}$ must have $W_T(a)>C$ and hence is not feasible. Any value of $a$ after $\hat{a}$ must have a larger value for $V(a)$, which we are trying to minimize. We can conclude that there exists a unique solution $a_{C}^*$ for the optimization problem and it is given by the unique solution to the equation $W_T(a)=C$.\eProof

\paragraph{\bf Proof of Proposition \ref{prop:opt_Gamma}.}
Throughout this proof we assume that $W_T(a_{\min})< C$. First, if $W_T(a_{\infty}^*)\leq C$, then $\Gamma=0$, and problems (\ref{eq:opt_uncap}) and (\ref{eq:opt_RSharing_modified_costs}) are identical and their solution is $a_{\infty}^*$. Theorem \ref{theorem:opt_a_cap} assures us that in this case, the solution to (\ref{eq:opt_RSharing_recall}) satisfies that $a_{C}^*=a_{\infty}^*$, which proves the desired result.

We turn to the case where $W_T(a_{\infty}^*)> C$. In this case, $\Gamma>0$ and Theorem \ref{theorem:opt_a_cap} assures us that $a_{\min}\neq a_{\infty}^*$ and that $a_{C}^*$ is the unique value of $a\in \mathcal{A}$ strictly between $a_{\min}$ and $a_{\infty}^*$ such that $W_T(a)=C$. In particular, $a_{C}^*$ must be an internal point in $\mathcal{A}$ and $W_T'(a_{C}^*)\neq 0$.

Next, we leverage a structural property inherent in $V(a)$. Recall that 
$$V(h_R,h_H,a) = h_R W_R(a) + \edit{\lambda} p_x(a)  h_T T +  h_H W_H(a),$$
and, therefore,
\begin{align*} V(h_R,h_H,a)+\Gamma W_T(a) &= h_R W_R(a)+\edit{\lambda}p_x(a)h_T T+h_HW_H(a)+\Gamma W_R(a)+\Gamma W_H(a)\cr
&=(h_R+\Gamma )W_R(a) + \edit{\lambda}p_x(a)h_TT + (h_H+\Gamma )W_H(a) = V(h_R + \Gamma ,h_H + \Gamma ,a).
\end{align*}
Namely, 
\begin{equation}\label{eq:v_equals_v_plus_w}
    V(h_R+\Gamma ,h_H+\Gamma ,a)=V(h_R,h_H,a)+\Gamma W_T(a).
\end{equation}
Taking the derivative of the right-hand side with respect to $a$ and using the definition of $\Gamma$, we obtain:
$$\left(V(h_R,h_H,a)+\Gamma W_T(a) \right)' = V'(h_R,h_H,a)-\frac{V'(h_R,h_H,a_{C}^*)}{W_T'(a_{C}^*)}W_T'(a).$$

Clearly, this derivative equals zero for $a=a_{C}^*$. By \eqref{eq:v_equals_v_plus_w}, this also means that the derivative of the left hand-side is zero for $a=a_{C}^*$. However, from the analysis in the proof of Proposition \ref{prop:opt_a_New2}, we know that $V'(h_R,h_H,a)$ (for \textit{any} $h_R,h_H>0$) can be zero at most once in $\mathbb{R}_+$. Moreover, if $V'(h_R,h_H,\tilde{a})=0$ for $\tilde{a}\in (0,\bar{A})$, then $\tilde{a}$ is a unique global minimum of $V(h_R,h_H,a)$ in $\mathcal{A}$. Therefore, $a_{C}^*$ is the unique solution to \eqref{eq:opt_RSharing_recall}, which concludes the proof. \eProof


\paragraph{\bf \edit{Proof of Proposition} \ref{prop:feasibility_shared_Mult}.}
\edit{This proof follows very similarly to that of Proposition~\ref{prop:feasibility_shared}. In the discussion of section \ref{subsec: The Multi-Type Feasibility Region}, we establish that the feasibility region of optimization problem \eqref{eq:opt_RSharing_Mult} is
\[
{\calC_{FR}^K} = \left\{(\vec{\lambda},C) \in \mathbb{R}_+^{K+1} :  \sum_{k=1}^K W_T^k(a_{\min}^k)\leq C \right\}.
\]
All that is left is to characterize $a_{\min}^k$ for $k=1,\ldots,K$. By Proposition \ref{prop:W_total_on_a}, if $\theta_H^k/\theta_R^k\leq 1$, then $W_T^k(a^k)$ is strictly decreasing in $a^k$. Thus, in this case, $a_{\min}^k=0$, which proves the first item.
Next, by Proposition \ref{prop:W_total_on_a}, if $\theta_H^k/\theta_R^k \geq 1+\Delta^k$, then $W_T^k(a^k)$ is strictly increasing in $a^k$. Thus, in this case, $a_{\min}^k=\bar{S}^k-x^k-T^k\theta_T^k$, which proves the second item.}

\edit{Finally, in the case where $1<\theta_H^k/\theta_R^k<1+\Delta^k$, by Proposition \ref{prop:W_total_on_a}, there exists $a_0^k>0$ such that $W_T^{k'}(a^k)<0$ for $a^k \in [0,a_0^k)$, $W_T^{k'}(a_0^k)=0$, and  $W_T^{k'}(a^k)>0$ for $a^k \in (a_0^k,\infty)$. If $a_0^k < \bar{S}^k-x^k-T^k\theta_T^k$, then $a_{\min}^k=a_0^k$. Otherwise, $\bar{S}^k-x^k-T^k\theta_T^k \in (0,a_0^k]$, and since $W_T^k(a^k)$ is strictly decreasing in this interval, we have $a_{\min}^k=\bar{S}^k-x^k-T^k\theta_T^k$ which completes the proof of the third item and the proposition.} 
\eProof

\paragraph{\bf \edit{Proof of Lemma \ref{lemma: multi type existence}}.}
\edit{If the feasibility region contains exactly one vector, then this vector is optimal. Otherwise, since the total workloads are continuous functions, the feasibility region contains an infinite number of vectors. We begin by proving that in this case, the feasibility region is a compact set, by proving it is bounded and closed. The set $\cal{C}_{FR}^K$ is bounded because 
$$\Vert \vec{a}-\vec{b} \Vert_2\leq \Vert \vec{a}\Vert_2+\Vert \vec{b}\Vert_2\leq 2K\max_k {\left\{\bar{A}^k\right\}}, \quad \forall a,b\in \cal{C}_{FR}^K.$$
Next, let $\vec{c}$ be a limit point of $\cal{C}_{FR}^K$. Assume by contradiction that $\vec{c}\notin \cal{C}_{FR}^K$. First, consider the case where there exists an entry $k$ such that $[\vec{c}]_k\notin [0,\bar{A}^k]$, meaning, it is outside of the hypercube $[0,\bar{A}^1]\times \ldots \times [0,\bar{A}^K]$. Clearly, there is a small enough neighbourhood of $\vec{c}$ with no points that belong to $\cal{C}_{FR}^K$, which contradicts the fact that $\vec{c}$ is a limit point. Therefore, $\vec{c}$ must be in the hypercube.}

\edit{We are left with the case where $\sum_{k=1}^K W_{T}^k(c^k) >C$, meaning that there exists $\delta>0$ such that $\sum_{k=1}^K W_{T}^k(c^k) =C+\delta$. Since $\{W_{T}^k\}$ are continuous functions, so is $\sum_{k=1}^K W_{T}^k(c^k)$ as a function from $\mathbb{R}_+^K$ to $\mathbb{R}_+$. Hence, there exists a small enough neighborhood of $\vec{c}$ such that for every point $\vec{a}$ in it we have $\sum_{k=1}^K W_{T}^k(a^k) >C+\delta/2$. Thus, no points in this neighborhood belong to $\cal{C}_{FR}^K$ which, again, is a contradiction. Hence, $\cal{C}_{FR}^K$ is closed.}

\edit{Since $\cal{C}_{FR}^K$ is compact and the cost function is continuous, by the Extreme Value Theorem the infimum is attained and there exists an optimal solution.} \eProof

\paragraph{\bf \edit{Proof of Lemma \ref{lemma: multi type optimality if in FR}}.}
\edit{By Proposition \ref{prop:opt_a_New2}, $[\vec{a}^*_\infty]_k$ is the unique minimizer of $V_k([\vec{a}]_k)$. Since $V(\vec{a})=\sum_{k=1}^K V_k([\vec{a}]_k)$, $\vec{a}^*_\infty $ is the unique minimizer of $V(a)$.} 
\eProof

\paragraph{\bf \edit{Proof of Theorem \ref{theorem: multi type sol char}}.}
\edit{For each type, we know from Propositions \ref{prop:W_total_on_a} and \ref{prop:opt_a_New2} and their proofs that $V_k$ and $W^k_T$ each can exhibit one of the following behaviours in the allowable threshold interval: (1) strictly decreasing, (2) strictly increasing or (3) strictly decreasing, attains a minimum and strictly increasing afterwards. We use the symbols $\searrow$, $\nearrow$, and $\searrow \nearrow$ to refer to these cases respectively. Thus, there are nine possible combinations of how $V_k$ and $W^k_T$ behave for a specific type. For example, there is a possibility that $V_k$ is strictly increasing while $W^k_T$ is strictly decreasing. We will use the notation $V_k \searrow W^k_T \nearrow$ for this case. We now prove that for each combination, it is impossible or necessarily sub-optimal to choose the threshold not between $a_{\min}^k$ and $a_{\infty,k}^*$.}

\begin{enumerate}
    \item \edit{$V_k \nearrow W^k_T \nearrow$: In this case $a_{\min}^k=a_{\infty,k}^*=0$. Any other choice of threshold would result in a higher cost and a higher workload and therefore is sub-optimal.}
    \item \edit{$V_k \nearrow W^k_T \searrow$: In this case $a_{\min}^k=\bar{A}^k$ and $a_{\infty,k}^*=0$ and therefore any choice of threshold has to be between the two.}
    \item \edit{$V_k \nearrow W^k_T \searrow\nearrow$: In this case $a_{\infty,k}^*=0$ and $a_{\min}^k$ is an interior point. A choice of a threshold to the right of $a_{\min}^k$ results in a higher cost and a higher workload than choosing, for example, $a_{\min}^k$, and therefore is sub-optimal.} 
    \item \edit{$V_k \searrow W^k_T \nearrow$: Similar to case 2.}
    \item \edit{$V_k \searrow W^k_T \searrow$: Similar to case 1.}
    \item \edit{$V_k \searrow W^k_T \searrow\nearrow$: Similar to case 3.}
    \item \edit{$V_k \searrow \nearrow W^k_T \nearrow$: In this case $a_{\min}^k=0$ and $a_{\infty,k}^*$ is an interior point. Choosing a threshold to the right of $a_{\infty,k}^*$ results in a higher cost and a higher workload than choosing, for example, $a_{\infty,k}^*$, and therefore is sub-optimal.} 
    \item \edit{$V_k \searrow \nearrow W^k_T \searrow$: Similar to case 7.}
    \item \edit{$V_k \searrow \nearrow W^k_T \searrow\nearrow$: In this case both $a_{\infty,k}^*$ and $a_{\min}^k$ are interior points. If they are equal we are done. If they are not, assume that $a_{\infty,k}^*<a_{\min}^k$. Choosing a threshold to the right of $a_{\min}^k$ results in a higher cost and a higher workload than choosing, for example, $a_{\min}^k$, since both $V_k$ and $W_T^k$ are strictly increasing to the right of $a_{\min}^k$, making this a sub-optimal choice. The same holds for choosing a threshold to the left of $a_{\infty,k}$. The case where $a_{\infty,k}^*>a_{\min}^k$ is similar.} 
\end{enumerate}

\edit{This concludes the proof of the first item in Theorem \ref{theorem: multi type sol char}. For the second item, assume by way of contradiction that the capacity constraint at $\vec{a}^*_C$ is inactive, i.e., $\sum_k W^k_T([\vec{a}^*_C]_k)<C$. 
Since we are under the assumption that $\vec{a}^*_\infty \notin \cal{C}_{FR}^K$, by the first item in Theorem \ref{theorem: multi type sol char}, there must be at least one type $\bar{k}$ for which $[\vec{a}^*_C]_{\bar{k}}$ is between $a_{\infty,\bar{k}}^*$ and $a_{\min}^{\bar{k}}$ but not equal to $a_{\infty,\bar{k}}^*$. 
In all of the nine combinations we considered for the behavior of $V_k$ and $W^k_T$, and as we show in the proof of Theorem \ref{theorem:opt_a_cap}, 
when we move from $a_{\min}^{\bar{k}}$ towards $a_{\infty,\bar{k}}^*$ the cost strictly decreases and the total workload strictly increases. 
Thus, given that $\sum_k W^k_T([\vec{a}^*_C]_k)<C$, we have some slack, and there exists a threshold between $[\vec{a}^*_C]_{\bar{k}}$ and $a_{\infty,\bar{k}}^*$ such that if we choose it instead of $[\vec{a}^*_C]_{\bar{k}}$ and leave all other entries of $\vec{a}^*_C$ the same we get a feasible solution with a strictly smaller cost than with the choice of $\vec{a}^*_C$. This contradicts the fact that $\vec{a}^*_C$ is a globally optimal solution. This concludes the proof of the second item in Theorem \ref{theorem: multi type sol char}.}

\edit{For the third item, recall that $E$ denotes the set of indices for which the corresponding entries of the optimal solution $\vec{a}^*_C$ are not equal to $a^k_{\min}$ or $a^*_{\infty,k}$. Denote by $\vec{a}^*_{C,E}$ the solution $\vec{a}^*_C$ restricted to the entries in $E$. Consider the following optimization problem:
\be\label{eq:opt_RSharing_Mult_E_c}\begin{split}
&\min_{\vec{a} \in \vec{\mathcal{A}}_{E} } V(\vec{a}) \\
&\mbox{s.t. }  
\sum_{k\in E} W_{T}^k(a^k) = C-\sum_{k\in E^c} W_{T}^k([\vec{a}^*_C]_k),
\end{split} \ee
namely, we fix the entries in $E^c$, and optimize over the rest.}
\begin{lemma}\label{lemma:a_restricted_is_opt}
    \edit{$\vec{a}^*_{C,E}$ is an optimal solution for the optimization problem (\ref{eq:opt_RSharing_Mult_E_c}).} 
\end{lemma}

\paragraph{\bf \edit{Proof of Lemma \ref{lemma:a_restricted_is_opt}}.}
\edit{Assume by way of contradiction that there exist $\vec{b} \in \vec{\mathcal{A}}_{E}$ for which $V(\vec{b})<V(\vec{a}^*_{C,E})$. Then if we take the elements of $\vec{a}^*_C$ in the indices that belong to $E^c$ with the corresponding elements of $b$ we get a feasible vector for the original optimization problem but with a lower cost than that of $\vec{a}^*_C$. This contradicts the fact that $\vec{a}^*_C$ is an optimal solution.} 
\eProof

\edit{Continuing with the proof of the third item in Theorem \ref{theorem: multi type sol char}, we now prove that $\vec{a}^*_{C,E}$ is regular. By the first item in Theorem \ref{theorem: multi type sol char} and the definition of $E$, the entries of $\vec{a}^*_{C,E}$ must be strictly between the corresponding $a^k_{\min}$ and $a^*_{\infty,k}$ and therefore must be interior points. Thus, all of the boundary constraints aside from the capacity constraint are inactive. $\vec{a}^*_{C,E}$ is a feasible solution, so all that is left is to verify is that at least one of the derivatives ${W_T^k}'([\vec{a}^*_{C,E^c}]_k)$ is not zero. But, by Proposition \ref{prop:W_total_on_a}, these derivatives can only be zero once, at the corresponding $a^k_{\min}$. By the definition of $E$, $[\vec{a}^*_{C,E}]_k\neq a^k_{\min}$, for all $k \in E$. Thus, we conclude that  $\vec{a}^*_{C,E}$ is regular.}


\edit{Now, $\vec{a}^*_{C,E}$ is an optimal solution for the optimization problem (\ref{eq:opt_RSharing_Mult_E_c}), and it is regular. In addition, all of the inequality constraints are inactive. Thus, by the KKT sufficient conditions for optimality (e.g., Proposition 3.3.1. in \cite{bertsekas1997nonlinear}), we obtain that there exists a unique Lagrange multiplier $\Gamma\geq 0$ such that ${V'}_k([\vec{a}^*_C]_k)+\Gamma {W^k_T}'([\vec{a}^*_C]_k)=0$, for all $k\in E$. Moreover, by the definition of $E$ and the properties of $\{V_k\}$ and $\{W_T^k\}$, none of these derivatives are zero. Thus, $\Gamma$ must be strictly positive.}

\edit{Lastly, we wish to prove that $\vec{a}^*_{C,E}$ is the unique optimal solution of the un-capacitated optimization problem $\min_{\vec{a} \in \vec{\mathcal{A}}_{E} } \sum_{k\in E}V_k\left(h_R+\Gamma,h_H+\Gamma,a_k\right)$, where $\Gamma>0$ is the previously considered unique Lagrange multiplier for which $\Gamma=-{V'}_k([\vec{a}^*_C]_k)/ {W^k_T}'([\vec{a}^*_C]_k)$ for all $k\in E$. The proof now follows the exact same steps as that of Proposition \ref{prop:opt_Gamma}. Namely, for each $k\in E$, $[\vec{a}^*_C]_k$ must be the unique minimum of $V_k\left(h_R+\Gamma,h_H+\Gamma,a_k\right)$ in the allowed interval. Thus, $\vec{a}^*_{C,E}$ is the unique minimizer of their sum.}\eProof

\vspace{0.1cm}
\section{\edit{Justification of Total Long-Run Average Cost}}
\label{subsec: Justification of Total Long-Run Average Cost}

\edit{Let $A(t)$ be a renewal process of incoming patients of a single type. 
The patients' health scores evolve according to the description in Section \ref{subsec: Stochastic Dynamics of the Individual Health Score}. We assume that patients are independent, and thus their health scores evolve according to independent copies of $B_R,Z,B_H$, indexed by $k$.
Specifically, let
\begin{align*}
    &\calB^{R}(k,t)=x+\sigma_RB^{R}(k,t)-\theta_R t, \cr
    &\tau_{R}(k,x,a)=\inf\{t\geq 0:\calB^{R}(k,t)=0~ \text{or}~ \calB^{R}(k,t)=a+x \}, \cr
    &\calB^{H}(k,t)=x+a+Z(k,x,a,T)+\sigma_HB^{H}(k,t)-\theta_H t, \cr
    & \tau_{H}(k,x,a,Z(k,x,a,T))=\inf\{t\geq 0:\calB^{H}(k,t)=0 \}.
\end{align*}
For simplicity, we write $\tau_{R}(k)=\tau_{R}(k,x,a)$ and $\tau_{H}(k,Z(k))=\tau_{H}(k,x,a,Z(k,x,a,T))$. Thus, the cost of the $k$-th patient is given by:}
$$
\edit{V_k(a)=h_R\tau_{R}(k)+1_{\{\calB^{R}(k,\tau_{R}(k))=a+x\}}\big( h_T T +h_H\tau_H(k,Z(k))\big),}
$$
\edit{where $\{V_k(a)\}$ are i.i.d.~with}
$$
\edit{\Ex[V_k(a)]=\Ex[V(a)]=h_R \Ex \left[\tau_{R}(x,a)   \right] 
+
\big(h_T T + h_H  p_x(a)\Ex \left[\tau_{H}(x,a,Z)   \right] \big).}
$$
\edit{The average cost at time $t$ is then defined as:}
$$
\edit{V(a,t):=\frac{1}{t} \sum_{k=1}^{A(t)}V_k(a).}
$$

\edit{By the Renewal-Reward Theorem (e.g., Equation (31) on page 82 in  \citealp{stirzaker2005stochastic}), the long run average cost is given by:}
$$
\edit{V(a)=\lim_{t \rightarrow \infty}V(a,t)=\lambda \Ex[V(a)]=\lambda \left ( h_R \Ex \left[\tau_{R}(x,a)   \right] 
+
\left(h_T T + h_H  p_x(a)\Ex \left[\tau_{H}(x,a,Z)   \right] \right)\right ),}
$$
\edit{justifying Equation \eqref{eq:obj_func_linear} as the system's total long-run average cost.}


\section{\edit{Data Requirements and Parameter Estimation}}
\label{sec: Parameter Estimation from Data}

\edit{We model the evolution of patients' health scores as negative-drift Brownian motions. However, instead of estimating the drift and variance directly, we can utilize station-level LOS data, which hospitals typically collect consistently and systematically. Specifically, we need data on the duration of remote and on-site hospitalizations for patients.}

\edit{Under our model, the LOS follows an inverse Gaussian distribution -- a distribution that has been widely used to model LOS in healthcare models for many years (see \cite{whitmore1975inverse} and the more recent \cite{hashimoto2023re}).
Thus, the required data includes patients' LOS at each hospitalization stage. From this data, we can estimate the parameters of the distribution, which correspond to the expectancy and variance of $\tau_R$ and $\tau_H$. This estimation can be accomplished using simple maximum-likelihood estimation.}

\edit{We then estimate the health scores at call-in/discharge and subsequently build estimates for $\theta_H$ and $\theta_R$. For patients who were called in, the necessary data includes their health scores before and after traveling, allowing us to estimate $\theta_T$ and $p_x(a)$.}



\section{\edit{Incorporating Quadratic Holding Costs}}
\label{sec: Incorporating Quadratic Holding Costs}

\edit{In this section, we explore the solution under a quadratic holding cost structure, which brings up the recovery variance of coefficient, $\sigma_1^2$ and $\sigma_2^2$, in Stations 1 and 2, respectively.} 

\edit{Recall that $\tau_R(x,a)$ and $\tau_H(x,a,Z)$ correspond to the (random) LOS in remote and on-site hospitalization, respectively. Consider the following holding cost functions:
\[
C_R(x,a) = h_R\left(\tau_R(x,a)\right)^2, ~~~ C_H(x,a,Z) = h_H\left(\tau_H(x,a,Z)\right)^2,
\]
for some $h_H,h_R>0$. Standard results for the hitting time of Brownian motions yield
\begin{align*}
\Ex\left[\tau_R(x,a)\right]^2 & = \frac{2}{\theta_R^2} \left( p_x(a) - \frac{\theta_R x}{\sigma_R^2} \right);\\
\Ex\left[\tau_H(x,a,Z)\right]^2 & = \left(\frac{x+a+T\theta_T}{\theta_H^3}\right)\sigma_H^2 + \left(\frac{x+a+T\theta_T}{\theta_H}\right)^2.
\end{align*}
Accordingly, the total long run average cost is:
\begin{equation}\label{eq:obj_func_quad}
\begin{split}
V(a) 
&=  
\lambda\left(h_R \Ex \left[\tau_{R}(x,a)   \right]^2 
+
\left(h_T T + h_H \Ex \left[\tau_{H}(x,a,Z)   \right]^2 \right) p_x(a)\right)
\\
&=
\lambda \left(\frac{2 h_R }{\theta_R^2} \left( p_x(a) - \frac{\theta_R x}{\sigma_R^2} \right) 
+
\left(h_T T + h_H \left(\left(\frac{x+a+T\theta_T}{\theta_H^3}\right)\sigma_H^2 + \left(\frac{x+a+T\theta_T}{\theta_H}\right)^2 \right)\right) p_x(a)\right), 
\end{split}
\end{equation}
which is equivalent to 
\[
V(a) 
=  \vartheta  + \delta a p_x(a) + \phi a^2 p_x(a) + \psi p_x(a),
\]
where 
\begin{align*}
\vartheta &= - 2\lambda \left(\frac{ h_R x}{\theta_R \sigma_R^2}\right), ~~~
\delta = \lambda h_H \left(\frac{\sigma_H^2}{\theta_H^3} + \frac{2\left(x+T\theta_T\right)}{\theta_H^2} \right), ~~~
\phi = \lambda\left(\frac{ h_H }{\theta_H^2}\right), \\
\psi &= \lambda \left( \frac{2 h_R }{\theta_R^2} + h_T T + h_H \left( \left(\frac{x+T\theta_T}{\theta_H^3} \right)\sigma_H^2 + \left(\frac{x+T\theta_T}{\theta_H} \right)^2 \right) \right).
\end{align*}}

\edit{Figure \ref{fig: optimal a and p - Quadratic costs} illustrates the solution structure when costs are quadratic when $\sigma_H=2$ (top) and $\sigma_H=8$. While the variability decreases the call-in thresholds and shrinks the interval where remote-hospitalization is effective, the results do not fundamentally change compared to our baseline model. This demonstrates the robustness of our suggested policy and insights.}  

\begin{figure}[h]
\caption{\edit{Optimal call-in threshold and call-in probability for quadratic costs as a function of travel time for different initial health scores. The parameters are $\theta_H=0.05$, $\theta_R=0.06$, $\theta_T=0.1$, $h_H =2.65$, $h_R =5.1$, $h_T =2$, $\gamma = -32$, $\lambda = 1$, $\sigma_R=1$, $\bar{S}=15$, in the top plots $\sigma_H =2$ and in the bottom plots $\sigma_H =8$.}}
\centering
\begin{subfigure}
\centering
\includegraphics[width=0.491\textwidth]{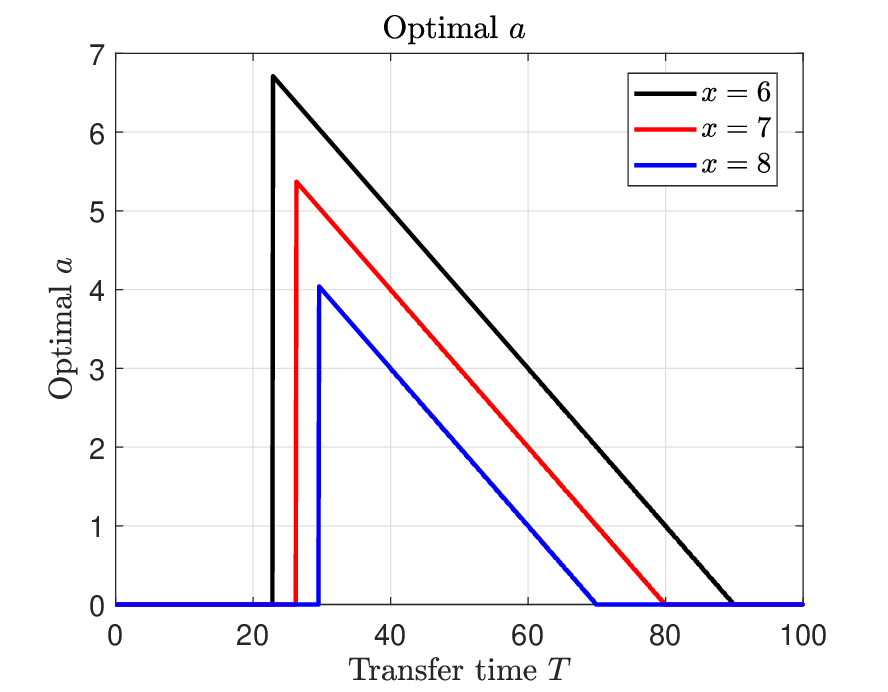}
\end{subfigure}
\centering
\begin{subfigure}
\centering
\includegraphics[width=0.491\textwidth]{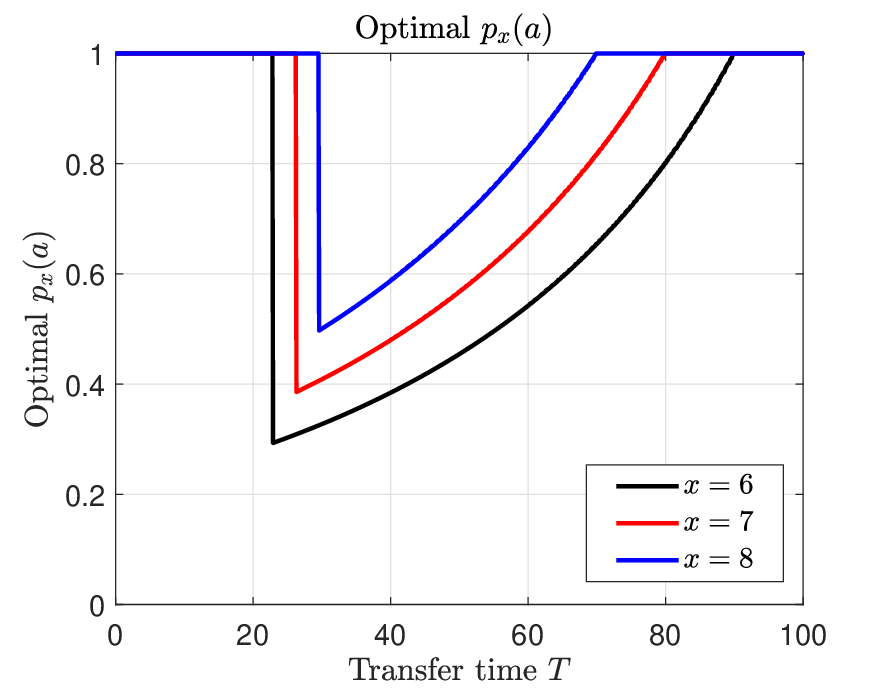}
\end{subfigure}\centering
\begin{subfigure}
\centering
\includegraphics[width=0.491\textwidth]{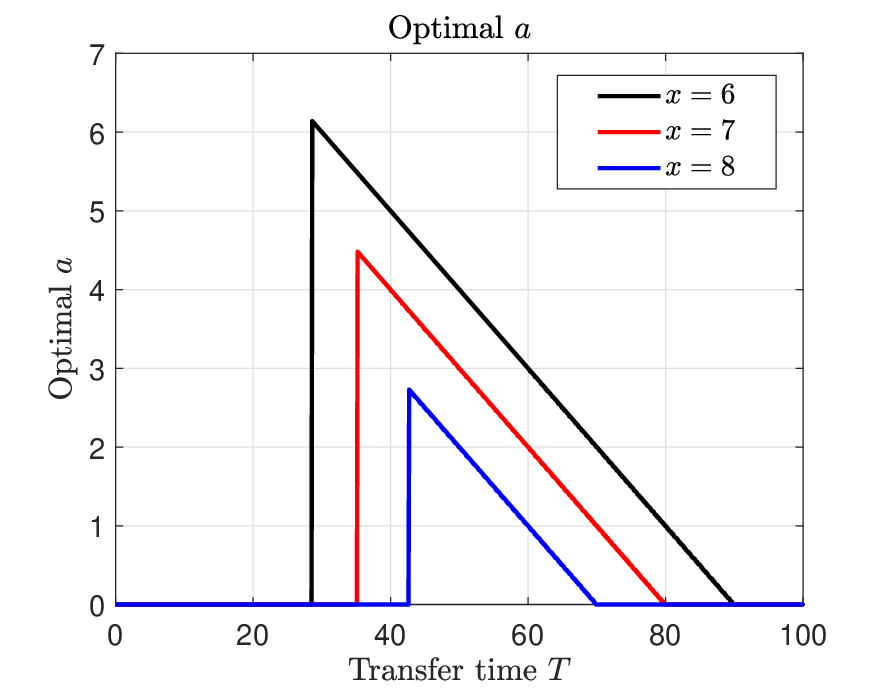}
\end{subfigure}
\centering
\begin{subfigure}
\centering
\includegraphics[width=0.491\textwidth]{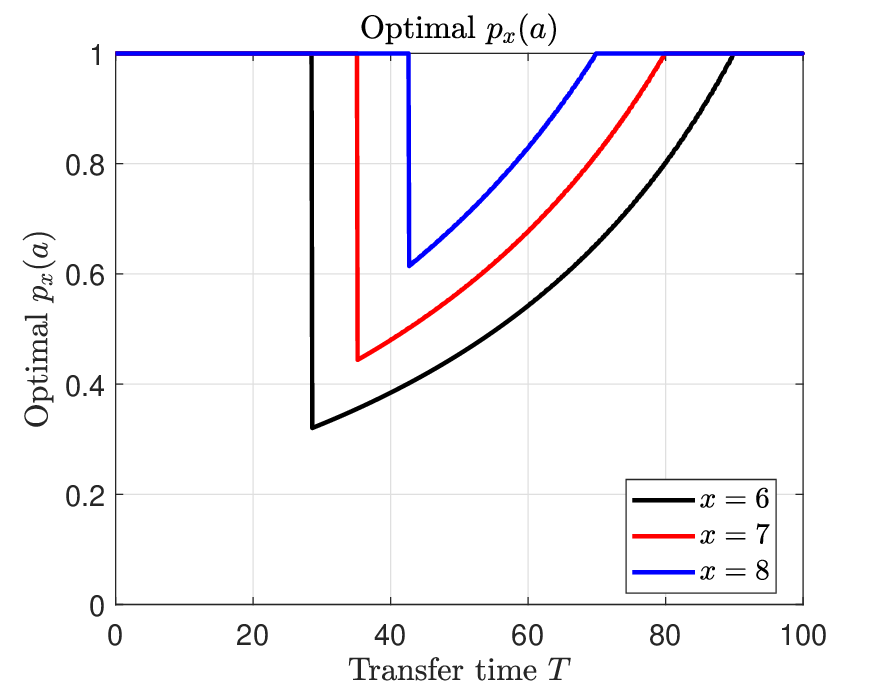}
\end{subfigure}
\label{fig: optimal a and p - Quadratic costs}
\vspace{-0.3in}
\end{figure}






\end{APPENDICES}

\end{document}